\documentclass[review]{elsarticle}

\usepackage{lineno,hyperref}
\usepackage{tabularx}
\newcommand\setrow[1]{\gdef\rowmac{#1}#1\ignorespaces}
\newcommand\clearrow{\global\let\rowmac\relax}
\clearrow
\usepackage[T1]{fontenc}
\usepackage[utf8]{inputenc}
\usepackage{amsmath}
\usepackage{adjustbox}
\usepackage{rotating}
\usepackage{natbib}
\usepackage{graphicx}
\usepackage{gensymb}
\usepackage{url}
\usepackage{mathtools}
\usepackage[varg]{txfonts}
\usepackage{textcomp}
\usepackage[super]{nth}
\usepackage{caption}
\usepackage{subcaption}
\usepackage{xcolor}
\usepackage[normalem]{ulem}

\usepackage{esvect}

\usepackage{siunitx}
\biboptions{authoryear}

\begin{document}

\begin{frontmatter}

\title{Coordinated Optical and Radar measurements of Low Velocity Meteors}

\author[uwopa,wiese]{Peter Brown}
\ead{pbrown@uwo.ca}
\author[uh]{Robert J. Weryk}

\address[uwopa]{Department of Physics and Astronomy, University of Western Ontario, London, Ontario, N6A 3K7, Canada}
\address[wiese]{Western Institute for Earth and Space Exploration, University of Western Ontario, London, Ontario, N6A 5B7, Canada}
\address[uh]{Institute for Astronomy, University of Hawaii, Honolulu HI, 96822, USA}

\begin{abstract}
To better estimate which luminous efficiency ($\tau$)  value is compatible with contemporary values of the ionization coefficient ($\beta$), we report a series of simultaneous optical and specular echo radar measurements of low speed (v < 20 km/s) meteors. We focus on the low speed population as secondary ionization is not relevant and the initial trail radii are small, minimizing  model assumptions  required to estimate electron line density. By using the large decrease in expected ionization coefficient at such low speeds, we attempt to better define the likely ratio of photon to electron production. This provides an estimate of the probable luminous efficiency, given that recent lab measurements of ionization efficiency agree with established theory \citep{jones1997, DeLuca2018} suggesting $\beta$ is more constrained than $\tau$.

Optical measurements were performed with two pairs of autonomously operated electron-multiplied charge coupled device cameras (EMCCDs) co-located with the multi-frequency Canadian Meteor Orbit Radar (CMOR) \citep{Brown2008}. Using the timing and geometry of individual meteors measured by both the radar and multi-station EMCCD systems, the portion of the optical lightcurve corresponding to each specular radar echo is measured and the received echo power used to estimate an electron line density.  A total of 1249 simultaneous EMCCD and radar meteors were identified from observations between 2017 – 2019 with 55 having in atmosphere speeds below 20 km/s. A subset of 36 events were analyzed in detail, with 29 having speed $<$ 20 km/s. These meteors had G-band magnitudes at the specular radar point between +4 and +7.7, with an average radiant power of 5W (assuming a 945 W power for a zero magnitude meteor). These correspond to a typical magnitude of +6.  Following the procedure in \citet{WerykBrown2013}, the ratio of {electron line density (q) to radiant power (I) } provides a direct estimate of the ionization coefficient ($\beta$) to luminous efficiency ($\tau$) ratio for each event. We find that $\beta$ / ${\tau}$ strongly correlates with radiant power. All our simultaneous meteors had asteroidal-like orbits and six were found to be probable iron meteoroids, representing 20\% of our slow $<$20 km/s sample. Luminous efficiency values averaged 0.6\% at low speed, ranging from $<$ 0.1\% to almost 30\%. No trend of luminous efficiency with speed was apparent, though a weak correlation between higher values of $\tau$ and radiant power may be present.  
\end{abstract}

\end{frontmatter}


\section{Introduction}

The measurement of meteoroid mass using either optical or radar observations of meteors requires knowledge of the amount of ablation energy partitioned into photon or electron production. The associated luminous efficiency ($\tau$) and ionization coefficient ($\beta$) are quantities which have historically been measured in the laboratory (eg. \citet{Slattery1967, Friichtenicht1968}) or estimated from meteor measurements. While the range in luminous efficiency estimates is very large \citep{Subasinghe2017}, recent laboratory measurements of the ionization efficiency \citep{DeLuca2018, Thomas2016} are in comparatively better agreement with the theoretical estimates from \cite{jones1997}.  Moreover, both measurements and theory suggest a rapid drop in ionization efficiency for speeds below 20 km/s. \par
Characterizing and understanding the meteoroid population encountering the Earth at low velocities (v < 20 km/s) has become increasingly important in recent years. In the last decade, new dynamical meteoroid models (eg. \citep{Nesvorny2010, Yang2015}) have predicted a large population of small, slow meteoroids originating from Jupiter-family comets which should dominate the mass influx to the Earth \citep{Carrillo-Sanchez2016}, in contrast to predictions from some earlier models \citep{Liou1995b}. These new models have produced estimates for the speed distribution as a function of meteoroid mass for such models \citep{Carrillo-Sanchez2016}, providing testable predictions. Similarly, recent studies (eg. \citet{Borovicka2005, Campbell-Brown2015}) have suggested that a significant fraction of low speed meteoroids may be iron in composition, with the fraction appearing to increase with decreasing mass \citep{Capek2019}. \par
Interpreting the physics behind the preceding examples depend on accurate estimation of meteoroid mass. This is a difficult problem, as meteoroid mass is always indirectly inferred from observations necessarily interpreted through the lens of model assumptions. At low speed, meteoroid mass becomes particularly uncertain as both the luminous efficiency (the fraction of total meteoroid kinetic energy which becomes radiation, $\tau$) and the ionization coefficient (the number of electrons produced per ablated atom, $\beta$) change at speeds below 20 km/s \citep{jones1997, jones2001, WerykBrown2013, DeLuca2018}. Dynamical mass estimates are complicated by the ubiquitous presence of fragmentation \citep{Subasinghe2016} which is significant at even very small meteoroid sizes \citep{Mathews2010}.\par
One approach to improve the accuracy of meteoroid mass estimates is to observe common meteor events using simultaneous (but independent) techniques. Comparing common mass estimates from different techniques allows for a sense of the global mass accuracy for individual meteors. In some cases, it is possible to combine measurements across techniques to better estimate energy conversion efficiencies. This approach has been most commonly employed through the fusion of optical and radar measurements of meteors, eg. \citet{Campbell-Brown2012}, through optical and infrasound \citep{Silber2015} as well as optical and LIDAR \citep{Klekociuk2005}, which have been combined to produce independent mass estimates.\par
Past studies using optical and radar measurements have focused either on comparing radar head echo (radial scattering) measurements to optical records \citep{Nishimura2001} or radar specular (transverse scattering) returns to optical signatures \citep{werykbrown2012}. The limitation of the former is that inference of meteoroid mass from radar head echo power returns depends on detailed knowledge of the plasma distribution in the immediate vicinity of the ablating meteoroid \citep{Marshall2017}, requiring models with many poorly constrained parameters \citep{Close2004}.

Specular radar returns, in contrast, are limited to probing the average electron line density in the vicinity of the first fresnel zone \citep{Ceplecha1998} and therefore can provide a snapshot only of mass loss in a short segment (~1-2 km) of trail. Furthermore, inference of the electron line density from transverse scattering depends on knowledge of the radial electron distribution \citep{Kaiser1952}, creating similar complications in interpretation as for the head echo case. However, for long wavelengths or low ablation heights, the initial trail radius may be small compared to the radar wavelength, minimizing the effects of attenuation \citep{Jones2005a}. In this limit, the reflected power is a relatively slowly varying function of the electron line density (particularly for echoes in the underdense regime), implying that the resulting electron-line density estimates are comparatively robust to modelling choices \citep{WerykBrown2013}. This also implies that estimates of electron line density are most accurate for lower altitude (and lower electron line density) echoes, a feature we exploit in this work.

The present study is an extension of the earlier work described in \citet{werykbrown2012} and \cite{WerykBrown2013}. That investigation focused on directly measuring the ratio of $\frac{\beta}{\tau}$ through comparison of specular echoes detected by the Canadian Meteor Orbit Radar (CMOR) and co-located (but manually operated) image intensified CCD cameras. Through adoption of a probable model for $\beta$, the corresponding values for $\tau$ were estimated. The major difference between this earlier work and the current study is the extension to much fainter optical meteors through the use of Electron-Multiplied Charge Coupled Devices (EMCCDs) as well as a focus on low speed events. In particular, the meteors reported in \citet{WerykBrown2013} primarily covered higher speeds; only half a dozen simultaneous radar-optical meteors had speeds below 20 km/s among the analysed sample of 129 events, which had an average speed of 44 km/s. Moreover, most of these low speed events were well past the transition regime between underdense and overdense where the uncertainty in electron line density is higher. The resulting luminous efficiency and $\frac{\beta}{\tau}$ in \citet{WerykBrown2013} are appropriate to speeds in excess of 20 km/s. 

Our goal here is to estimate $\frac{\beta}{\tau}$ in the speed interval 10\textless v \textless 20 km/s for the population of echoes which are in the underdense regime and hence fainter than the optical meteors reported by \citet{WerykBrown2013}. This is the range where we expect the most accurate estimates of electron line density to be possible.  We make use of the recent improved laboratory measurements for $\beta$ reported by \citet{DeLuca2018} to provide a means to directly estimate luminous efficiency and hence meteoroid mass at low speeds.

\section{Data Collection}
\subsection{Overview}
Our simultaneous optical-radar meteor observations were made between mid-2017 and mid-2019, and the data reduction process and methodology is similar to our previous work \citep{werykbrown2012, WerykBrown2013}.  Here we provide an overview of the key procedures used for analysis, highlighting the few differences from the earlier works, particularly with respect to optical measurements which were made with a new set of cameras.

The radar meteor data was collected by the Canadian Meteor Orbit Radar (CMOR) \citep{Jones2005, Brown2008}, while optical measurements were made using two new pairs of EMCCD cameras installed as a new  instrument suite as part of the Canadian Automated Meteor Observatory (CAMO) \citep{Weryk2013}. CAMO consists of two identical automated optical stations separated by 45 km, allowing optical triangulation of commonly observed meteor events. One of the CAMO stations is also co-located with the CMOR site; hence direct, common radar-optical measurements are possible.

CAMO runs automatically when clear, dark conditions are present. CAMO originally consisted of two distinct image intensified instrument suites. One was a mirror tracking system consisting of a wide-field intensified finder camera and a cued narrow field intensified camera attached to a telescope which viewed a pair of mirrors that tracked each meteor in real time. This system is collectively termed the "guided" system. Its design purpose was high temporal (10 ms) and spatial (3m) resolution studies of faint (+5) meteors. In addition to the guided system, is a wide-field fixed camera designed for population and flux studies to fainter meteor magnitudes (+6.5), termed the influx system. 

In 2016, four EMCCD cameras (two at each site) were installed at CAMO. This third optical system is optimized for population studies of faint, slow meteors, extending the sensitivity range of the influx system. As this is a new system for CAMO, we provide a detailed summary of the hardware and detection characteristics in section \ref{section:optical}. 

EMCCD camera data were collected at each site and analysed to isolate individual meteor events. Commonly observed events are correlated after each nightly run, and trajectory solutions computed. Any of these optical meteor trajectories which were within $\pm$ 3$^\circ$ of the specular point as seen from CMOR were flagged and the raw radar data centered around the time of the event (with a buffer of $\pm$ 5 sec) extracted and saved. The raw radar returns were then manually examined to search for echoes which had times within two seconds of the optical time estimate and interferometric locations within 3 degrees of the optically estimated specular point. This provided an initial filter to identify potential common optical-radar echoes. 

After this initial coarse filter, only two station optical solutions having average speeds below 20 km/s were selected for more detailed analysis. An optical meteor and radar echo were then positively associated as a common event if the following conditions were met:

\begin{enumerate}
    \item The interferometry direction was within 1.5$^\circ$ of the optical trail as measured from the EMCCD cameras at the CMOR site.
    \item Radar range of the echo and the range computed from two station optical solutions for the specular point agreed to <0.5 km.
    \item Timing of the radar peak power point and the frame containing two station optical solution for specular point are within 0.1 sec.
    \item Two station optical solution must have an angle between the two observation planes larger than 10 degrees.
\end{enumerate}

Using all automatically detected and computed EMCCD two station events collected between Oct 10, 2018 and May 10, 2019 (when the camera pointings were fixed and had common lenses and hence sensitivities) a total of 1249 possible optical events were selected in the first coarse filter as having a portion of their path within $3^{\circ}$ of the specular point. Of these, 96 had average optical speeds $<$ 20 km/s, but only 14 passed all remaining criteria given above. In addition to these 14, a further 15 events (collected between July 1, 2017 and Oct 9, 2018) were added which also passed the correlation criteria above. These earlier events were collected at a time when the EMCCD systems were still in an engineering testing mode with some lens changes and minor changes made to the pointing directions. While the collecting characteristics were not identical in this period, the long collection time significantly improved number statistics at low speeds and these events are added to our study data set here as they are clearly common with echoes detected by CMOR. Our analysis consists of these 29 low speed events (to which we also added 7 higher speed events to provide some context for the lower speed measurements) for a total of 36 radar-optical low speed meteors examined in detail for our study. \par

\subsection{Radar Measurements}

\begin{figure}
  \includegraphics[width=\linewidth]{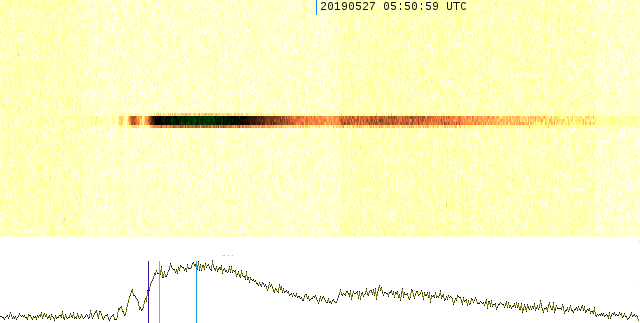}
  \caption{A raw radar range-time-intensity plot for the radar-optical meteor event detected on May 27, 2019 at 05:50:59 UTC. The top plot shows time on {the }abscissa - the entire axis as shown is 1.2 sec in duration. The ordinate is range from the radar running from 15-255 km. The brightness per pixel corresponds to the power received per sampled range gate. The echo at the main site is visible in the top plot - a horizontal slice of this echo power (in linear units) on the same time base is shown as the lower line plot. The vertical purple line is the estimated specular point while the {light } vertical blue line is the peak power point. The peak power point is the reference power level used to compute the electron line density {representing the full power scattered from the trail after the meteor has fully traversed the first fresnel zone. This power is corrected for diffusion back to the specular point. The dark blue line shows the location of the absolute peak power}.}
  \label{fig:radecho}
\end{figure}

Our specular radar measurements were made with the 29.85 MHz system of the Canadian Meteor Orbit Radar (CMOR). CMOR is a triple frequency radar operating at 17.45, 29.85 and 38.15 MHz having interferometric capability at each frequency. The 29.85 MHz system also has five additional remote sites which make common observations and permit 4000-5000 meteoroid orbits per day to be measured. Here we use only the data from the 29.85 MHz system, with focus on the amplitude, range, and interferometry solution from the main site. We found that most of the low speed common optical events did not have measureable radar orbits in the normal CMOR orbit pipeline, consistent with our observation that most of the common echoes were near the radar detection limit. Hence we restrict our radar analysis to the received echo power at the main site. An example of such a radar echo record for a simultaneously observed optical meteor is shown in Figure \ref{fig:radecho}. Details of the hardware, software, calibration and analysis pipeline are given in \citet{Webster2004, Jones2005, Brown2008, werykbrown2012, WerykBrown2013}.\par

The basic configuration for the radar is listed in Table \ref{tab:cmor}. The direction to each detected echo is found by interferometry using the phase differences (in pairs) between five antennas configured in a cross pattern as described in \citet{jones1998}. The interferometric accuracy is $\approx$ 0.8$^\circ$\citep{werykbrown2012}. Timing is GPS conditioned and absolute times per recorded pulse are accurate to much better than the EMCCD interframe time of 30ms. \par
The main difference in radar data compared to the earlier work of \citet{werykbrown2012, WerykBrown2013} is the higher transmit power (15 kW now versus 6 kW originally). As well, since raw radar returns are saved across all range gates for these echoes, the absolute range accuracy is higher than the range gate sampling interval (3 km) as the pulse shape is fit across each echo to isolate the peak range with an accuracy of 0.3 km. Since the optical field of view is limited to a comparatively small fraction of the whole sky, the limiting sensitivity for EMCCD optical detections is below CMORs absolute detection limit as shown in Table \ref{tab:cmor}. 

\begin{table*}[t] 
	\caption{Experimental configuration of CMOR for simultaneous radar-optical measurements. Here the total receiver and transmit gain in the direction of the fixed field of view EMCCD cameras is also shown, together with the equivalent limiting radar meteor magnitude in that direction.}
	\label{tab:cmor} 
	\centering 
	
	\begin{tabular}{l l} 
	\hline\hline 
	
Quantity  & Description \\ 
	\hline 
	Location & 43.264$\degree$N, 80.772$\degree$W, 324m (WGS-84)\\
Frequency & 29.85 MHz \\
Pulse duration & 75 $\micro$sec\\
Pulse Repetition Frequency & 532 Hz \\
Range sampling interval & 15 - 255 km \\
Peak Transmitter power & 15 kW \\
Range accuracy & $<$ 0.3 km \\
Total gain in FOV & 2.3 - 8.9 dBi \\
Limiting equivalent radar meteor magnitude in FOV & +5.7 - +7.5 \\

	\hline 
	\end{tabular}
\end{table*}

\subsection{Optical data}\label{section:optical}

\begin{figure}
  \includegraphics[width=\linewidth]{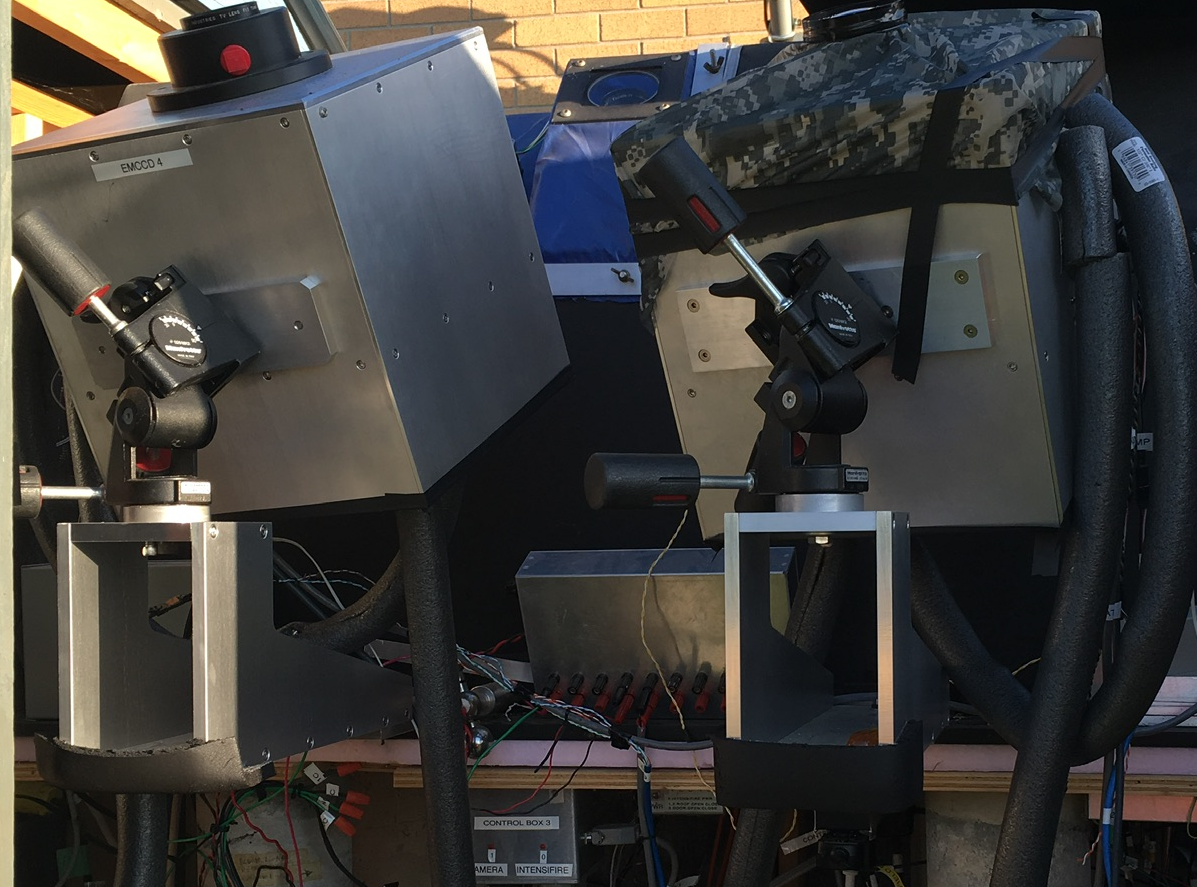}
  \caption{One pair of EMCCD cameras in-situ within the CAMO shed at the Elginfield site.}
  \label{fig:cams}
\end{figure}

Optical measurements used four N\"{u}v\"{u} HN\"{u}1024 EMCCD cameras \footnote{\url{http://www.nuvucameras.com/fr/files/2019/05/NUVUCAMERAS_HNu1024.pdf}}, two at each CAMO site. The cameras are Peltier cooled to -60$^\circ$C and thermally regulated via liquid cooling. They are pointed in fixed directions each with a 15x15 degree field of view. Figure \ref{fig:cams} shows a pair of cameras in situ. The cameras are GPS synchronized at the image stage and have absolute frame time accuracy better than 1 ms. They have a midband spectral response, broadly comparable to the Gaia, G-band \citep{Jordi2010}, and we use G-band magnitudes throughout our analysis. The quantum efficiency of each camera is in excess of 90\% between 500-700 nm. More details of the cameras are summarized in Table \ref{tab:EMCCDspecs}.

Each CAMO site has two EMCCDs which are paired to observe common atmospheric volumes.  The pair optimized for lower height events (cameras F) have a collecting area that is fairly flat with height (varying by less than  a factor of two from 80-140 km altitude) versus the other pair (cameras G) which have a more pronounced atmospheric collecting area maximum centered near 105 km. However, both pairs have significant collecting areas above 70 km with cameras F being slightly more sensitive than cameras G due to lower average ranges to events (higher pointing elevations). The limiting meteor detection sensitivity is near magnitude +8, though portions of the lightcurve approaching magnitude +9 are measurable above the EMCCD background for some meteors. This is an order of magnitude smaller mass than than the +5 limiting optical sensitivity of  \citet{WerykBrown2013}, and approaches or exceeds the radar sensitivity (which is about +7.5 equivalent magnitude) in the EMCCD field of view as seen from CMOR (Table \ref{tab:cmor}). 

The EMCCD image stream per camera is automatically searched using a modified cluster (blob) detector as a first stage for detection \citep{gural2016}. A matched filter is applied in a second stage for automated refined astrometry and photometry estimations. Camera events are then correlated across sites using the EVCORR algorithm as part of the ASGARD detection package \citep{Weryk2007} to produce trajectory and orbit solutions. Orbit counts per camera pair approach one per minute under good sky conditions. For this work, the automated solutions are used only as an initial filter for probable common optical-radar events. Once common events are identified and manually verified, all subsequent astrometric and photometric measurements are performed manually using the software METAL with its methods described by \citet{WerykBrown2013}.

\begin{table}
    \caption{Details of EMCCD. Pointing directions to the center of each field of view are given in degrees from the zenith ($\theta$) and azimuth ($\varphi$) measured E of N. }
    \centering
    \begin{tabular}{l l}
    \hline
    & Specification\\
    \hline
    CMOR Site [01] & 43.264$\degree$N, 80.772$\degree$W, 324m (WGS-84)\\
    Elginfield Observatory [02] & 43.194$\degree$N, 81.316$\degree$W, 319m ASL \\
    Pointing 1F/1G ($\theta , \varphi$) & 25$^\circ$, 323$^\circ$ / 43$^\circ$, 330$^\circ$ \\
    Sensor & Teledyne e2v CCD201-20\\
    Pixels & 1024x1024 @ 13$\mu$m\\
    Digitization & (14) / 16-bit\\
    Framerate (1x1/2x2) & 16.7/32.7 fps \\
    Lens & Nikkor 50 mm f/1.2 \\
    Field-of-View & 14.7$\degree$x14.7$\degree$ \\
    Meteor Peak G-Magnitude Limit & 8.0 \\
    Elginfield cameras F/G photometric offset & -16.29$\pm$0.33 / -16.01$\pm$0.35\\
    CMOR Site cameras F/G photometric offset & -16.89$\pm$0.06 / -16.62 $\pm$ 0.2\\
    \hline
    \end{tabular}
    \label{tab:EMCCDspecs}
\end{table}

\begin{figure}
  \includegraphics[width=\linewidth]{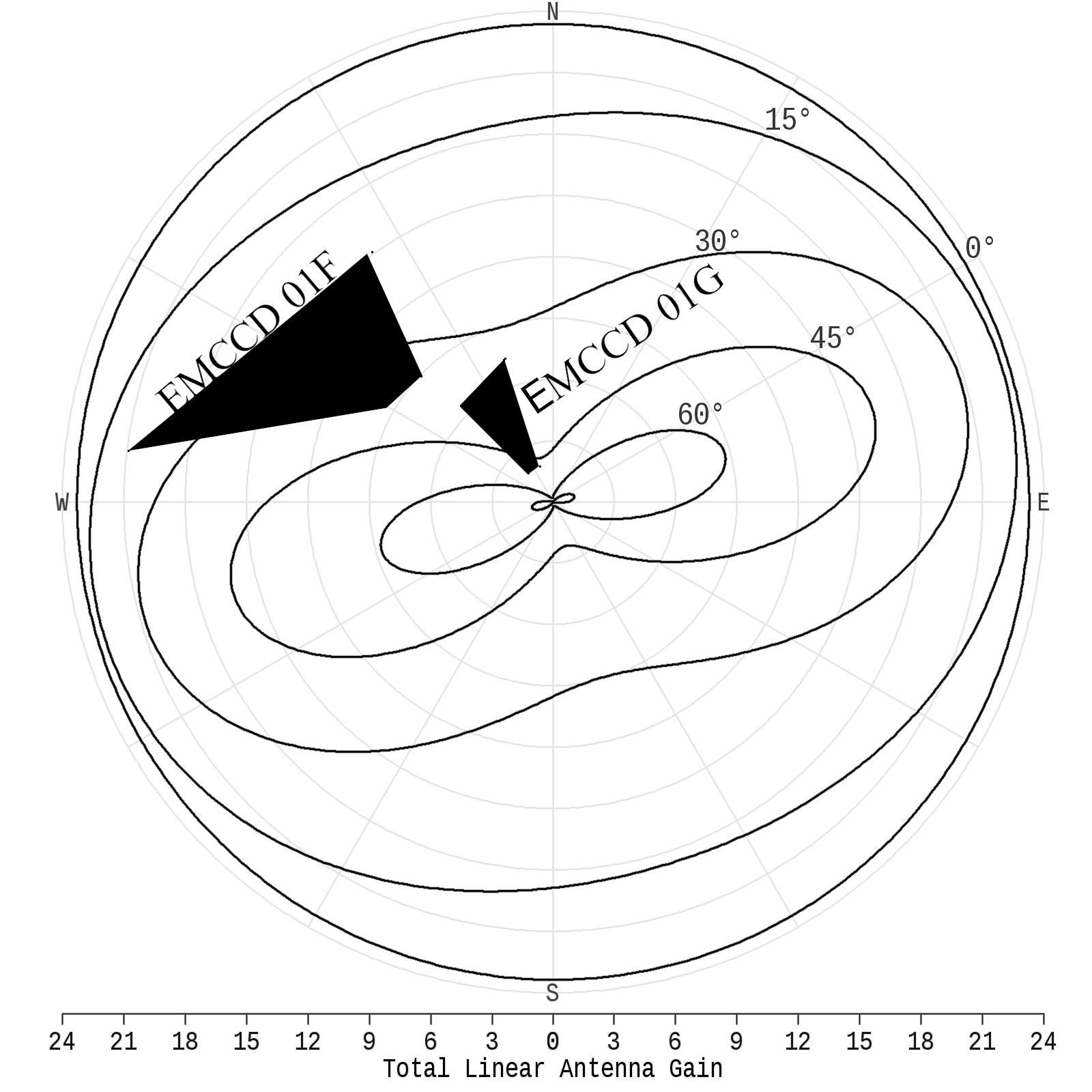}
  \caption{Field of view (FOV) of EMCCD cameras at the CMOR site projected on the sky with radar gain pattern overlay. The angular units show zenith distance as a function of azimuth with the horizon being at the center and the zenith around the edge of the circle.}
  \label{fig:betatauall}
\end{figure}

\begin{figure}
  \includegraphics[width=\linewidth]{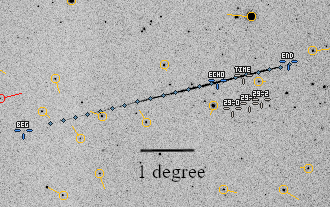}
  \caption{This meteor (20190527:055059) was detected by EMCCD 01F at the CMOR site and shows the full optical path from first detection (begin - in this case first seen from the other station) until final light (end). The individual blue symbols along the trajectory represent the manual astrometry picks per frame - this event lasted a total 20 EMCCD frames. The position marked "echo" is the two station optical solution estimate of the specular point - i.e. the location on the trail at $90^{\circ}$ from the apparent radiant as seen from the radar site. The time mark is the EMCCD time extracted from the radar when the echo specular point was reached. The ``29-0'' mark shows the estimated interferometric location of the echo as seen from the main site.  The ``29-1'' and ``29-2'' marks represent the apparent location on the plane of the sky where the three station radar solution places the apparent path of the radar trajectory. Here an offset of $0.5^{\circ}$ between the optical and radar path is apparent, likely due to phase offsets at the main radar site. The timing and echo points are concordant within two video frames or 0.06 sec while the 29-0 pick (the single station interferometric estimate of the echo location in the sky) is $0.5^{\circ}$ offset from the echo point, but parallel to the optical path. In this instance we have confidence from three independent measurements that the echo power is being produced along the lightcurve between the echo and time points, corresponding to an average absolute magnitude of +4.7 at the specular scattering point.}
  \label{fig:betatauall}
\end{figure}

\section{Methodology}\label{section:methods}
The core of our analysis leverages the fact that the ratio $\beta$ to $\tau$ can be estimated directly from simultaneous radar-optical meteor events. {From the fundamental physical theory of meteor ablation, it is assumed that the instantaneous radiant power, I, is proportional to a fraction $\tau$ of the kinetic energy of mass-loss via \citep{Ceplecha1998}

\begin{equation}
I= \frac{\tau v^2}{2} \frac{dm}{dt}
\label{eqn:tau}
\end{equation}

where we have ignored the contribution due to deceleration.\par
Similarly, the number of free electrons produced per unit trail length, $q$, is commonly assumed to be proportional to a constant, $\beta$, times the mass loss rate \citep{Ceplecha1998} giving

\begin{equation}
q= \frac{\beta}{\mu v} \frac{dm}{dt}
\label{eqn:beta}
\end{equation}
where $\mu$ is the average atomic mass of an ablated meteoroid atom, $v$ is the speed of the meteoroid, $q$ is the electron line density, and $I$ is the radiant power calibrated to our bandpass. We assume that $\mu$ is 24 amu consistent with a chondritic composition \citep{jarosewich1990}.\par
Dividing Eq. \ref{eqn:beta} by \ref{eqn:tau} we can then relate the two constants, $\tau$ and $\beta$, through purely observed quantities as \citep{Weryk2013a}
\begin{equation}
\frac{\beta}{\tau} = \frac{\mu v^3 q}{2 I}.
\label{eqn:betatau}
\end{equation}
}
 \par
We compute the average velocity, $v$, from the two station optical solution using the Monte Carlo lag-weighting technique of \cite{Vida2019a}. The uncertainty in speed is taken to be the difference in speed between stations. The initial speed computed from this approach is also used to compute the meteoroid orbits (see Table \ref{tab:orbits}). In most cases, even for our lowest speed events, the difference between the top of atmosphere initial speed and the average speed is found to be less than 1 km/s. \par
The radiant power per frame is computed by summing all the light from the meteor trail from the leading edge of the meteor back to the point of the leading edge from the previous frame after background subtraction. All photometry is calibrated to the Gaia G bandpass, with an assumed zero magnitude meteor having a radiant power of 945 W following \cite{Brown2017}. {This value is computed assuming meteor spectra can be fit as a 4500K equivalent blackbody, an approximation found to be appropriate for slower meteors \citep{Borovicka1993, Borovicka1994a, Ceplecha1998}. } For each meteor and each camera, a separate manual stellar photometric calibration was performed assuming the log-sum-pixel intensity ($P_i$) of a source is related to the G magnitude via:
\begin{equation}
\mathrm{G = C-2.5 log_{10}\sum P_i}. 
\label{eqn:linedensity}
\end{equation}

\begin{figure}
  \includegraphics[width=\linewidth]{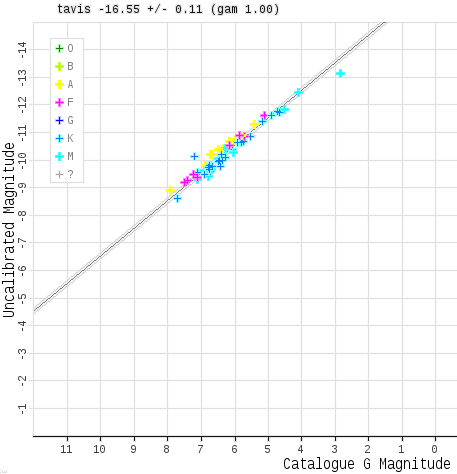}
  \caption{An example of a stellar photometric calibration for the CMOR EMCCD 'G' camera for the May 27, 055058 UTC event. The log sum pixel is shown on the ordinate and the calibrated G-band stellar magnitude is shown on the abscissa. For each star used in the calibration it's spectral type is shown {as colored symbols which follow the legend inset. }The zero point calibration value found through linear regression of the individual points (-16.55) plus its confidence bound (0.11) is also shown.}
  \label{fig:photocal}
\end{figure}

Here the zero point calibration C was found to be fairly constant over the two year period having a standard deviation among cameras for all events of between $0.06$ and $0.35$ magnitudes excluding the early intervals when engineering configurations were used, as shown in Table~\ref{tab:EMCCDspecs}. A sample stellar photometric calibration is shown in Figure \ref{fig:photocal}. The meteor magnitude uncertainties per frame are the combined uncertainties from the stellar calibration zero point uncertainty and photon counting statistical uncertainty. \par
The final term to be measured in Eq.~\ref{eqn:betatau} is the electron line density, $q$. From the calibrated echo power received from the meteor as measured at the receivers ($P_r$), we can relate the electron line density at our frequency via { \citep{Kaiser1952, Ceplecha1998}}:

\begin{equation}
\mathrm{g(q,a)} = \sqrt {\frac{P_r 32 \pi^4R^3}{P_T G_T G_r \lambda^3}}
\label{eqn:linedensity}
\end{equation}

The terms on the right hand side are all known or directly measured and include $R$, the range to the meteor echo, $P_T$ the transmit power, $G_T$ and $G_r$ the antenna gain in the direction of the echo for the transmit and receive antennas respectively, and $\lambda$, the radar wavelength. From these quantities, the reflection coefficient, $g$, of the trail can be determined. Here $g$ represents the fraction of the incident electric field reflected by the trail, a value which can range from 0 to in excess of two \citep{Poulter1977} due to polarisation effects. The reflection coefficient is a function of both the electron line density and its radial distribution including the trail radius, a.\par
Estimating $q$ from $g$ requires a scattering model which depends on the orientation of the trail relative to the polarization direction of the radar pulse, a quantity which is known provided the trail direction is known (which is the case for all our optically measured events). It also depends on the radius, $a$, of the cylindrical trail column, the electron line density and the radial distribution of electrons in the trail. For the latter, a Gaussian distribution has traditionally been assumed \citep{Kaiser1952} a choice which is supported by more recent simulations \citep{jones1995}. We estimate the initial radius as a function of height using the multi-frequency measurements from CMOR reported by \cite{Jones2005a}. We then compute the value of q needed to produce the observed $g$ given the observed specular echo height (and hence trail radius corrected for ambi-polar diffusion) using the full wave scattering model of \citet{Poulter1977} as implemented through look-up tables computed by \citet{WerykBrown2013} for CMOR's wavelength.\par

\section{Results and Discussion}
Table ~\ref{tab:orbits} summarizes the measured velocities, orbits, and magnitudes at the specular point and corresponding electron line density for all our simultaneous radar-optical meteors. Also shown are the meteor begin and end heights from optical data as well as the height of the radar echo based on the interferometry and range. Note in three cases the radar height is slightly (\textless 0.5 km) below the end height. This reflects the order of the uncertainty in the radar height; in these cases the specular point is taken to be at the end of the optical trail. The lightcurves for all events are shown in \ref{appendix:lightcurves}.\par
Using the data from Table ~\ref{tab:orbits} we compute the ratio of $\beta$ / ${\tau}$ per event and its uncertainty, which reflects only our uncertainty in $v$ and $I$ - this is shown in Table ~\ref{tab:betatau}.  Figure ~\ref{fig:betatauall} shows the resulting values of $\beta$ / ${\tau}$ as a function of speed.\par
\begin{figure}
  \includegraphics[width=\linewidth]{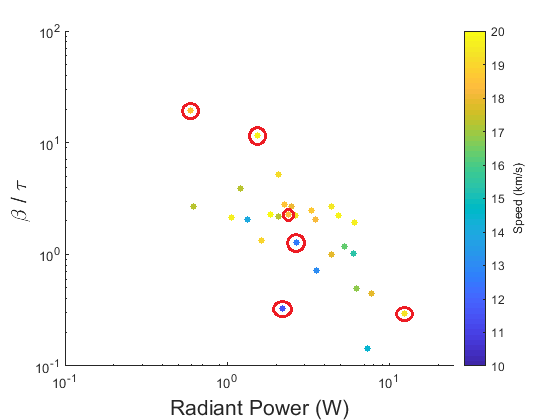}
  \caption{$\beta$ / ${\tau}$ as a function of radiant intensity color coded by speed for low velocity ($<$ 20km/s) events only. {Iron candidates are circled in red.}}
  \label{fig:beta-tau-int}
\end{figure}
\begin{figure}
  \includegraphics[width=\linewidth]{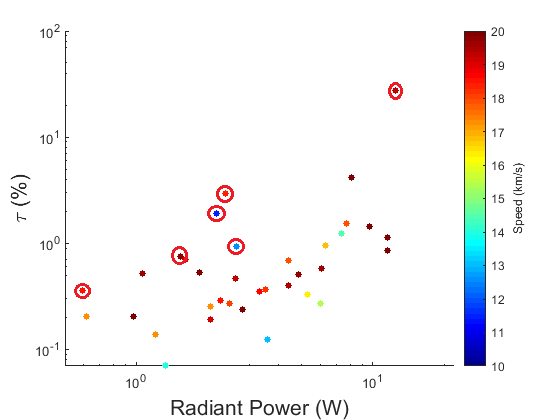}
  \caption{${\tau}$ as a function of radiant intensity colour coded by speed for slow ($<$ 20km/s) events only. Here $\beta$ is assumed to be the chondritic mean value reported in \citet{WerykBrown2013}, except for our six probable iron meteoroids which use the iron $\beta$ from \citet{DeLuca2018}.~{ Iron candidates are circled in red.}}
  \label{fig:tauvsI}
\end{figure}

Our measured sample shows scatter in $\beta$ / ${\tau}$ for a particular speed, but much of this variance is correlated with radiant intensity. In particular, the faintest events all have the largest $\beta$ / ${\tau}$ as shown in Figure~\ref{fig:beta-tau-int}. The most recent lab experiments show no evidence for a variation of $\beta$ with mass at low speeds \citep{DeLuca2018}. Our results suggest that $\tau$ varies with radiant intensity as shown in Figure \ref{fig:tauvsI} . This is consistent with the trend found in \cite{WerykBrown2013}, although their data was for brighter and faster meteors. It is the opposite trend found by \citet{Subasinghe2018} and \citet{Capek2019}, though in both cases they indicate that the trend is weak. \par
Examination of Table~\ref{tab:orbits} shows that all of our simultaneous radar-optical meteors have orbits with T$_j$ > 3. This suggests that our sample is dominated by asteroidal meteoroids. Among our sample of 36, six events each have abrupt onset lightcurves with low ($<$20 km/s) speed, and are highlighted in the tables in bold. Such sudden onset lightcurves have been suggested as a likely feature of iron meteoroids by \cite{Capek2019}. In their work, they identified probable iron meteoroids using a combination of spectral information and begin height/trail length criteria. In particular, through modelling, \citet{capekboro2017} suggested that pure iron meteoroids should also begin ablation at lower heights than regular stony meteoroids and should show shorter trail lengths on average.\par
\begin{figure}
  \includegraphics[width=\linewidth]{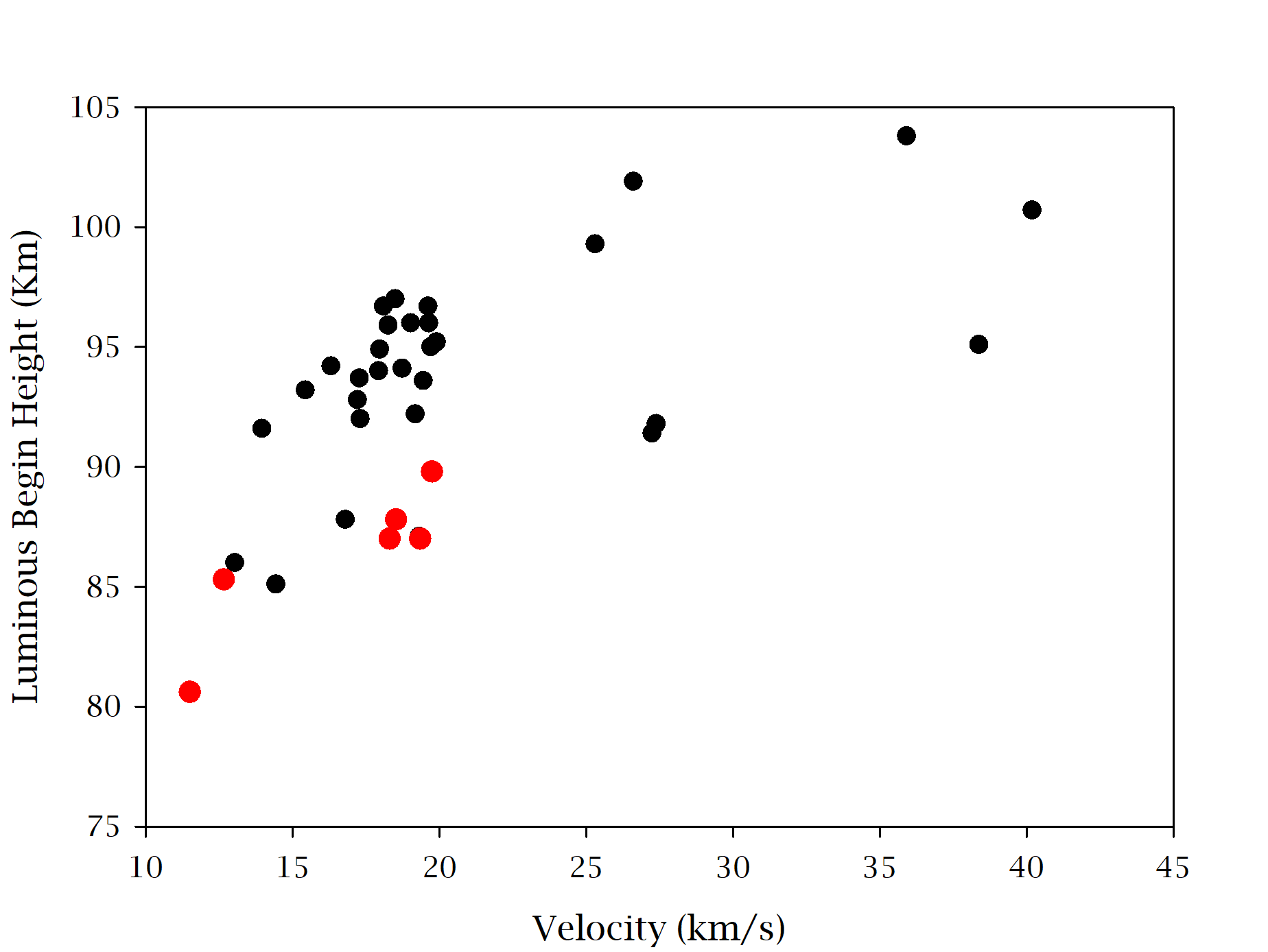}
  \caption{Begin height as a function of initial speed for all events in our dataset. As we required complete lightcurves from all events these represent the true initial heights - ie. the height at which luminosity first exceeds the instrument limit. Red symbols represent the six events which display sudden onset lightcurves.}
  \label{fig:ht-speed}
\end{figure}
\begin{figure}
  \includegraphics[width=\linewidth]{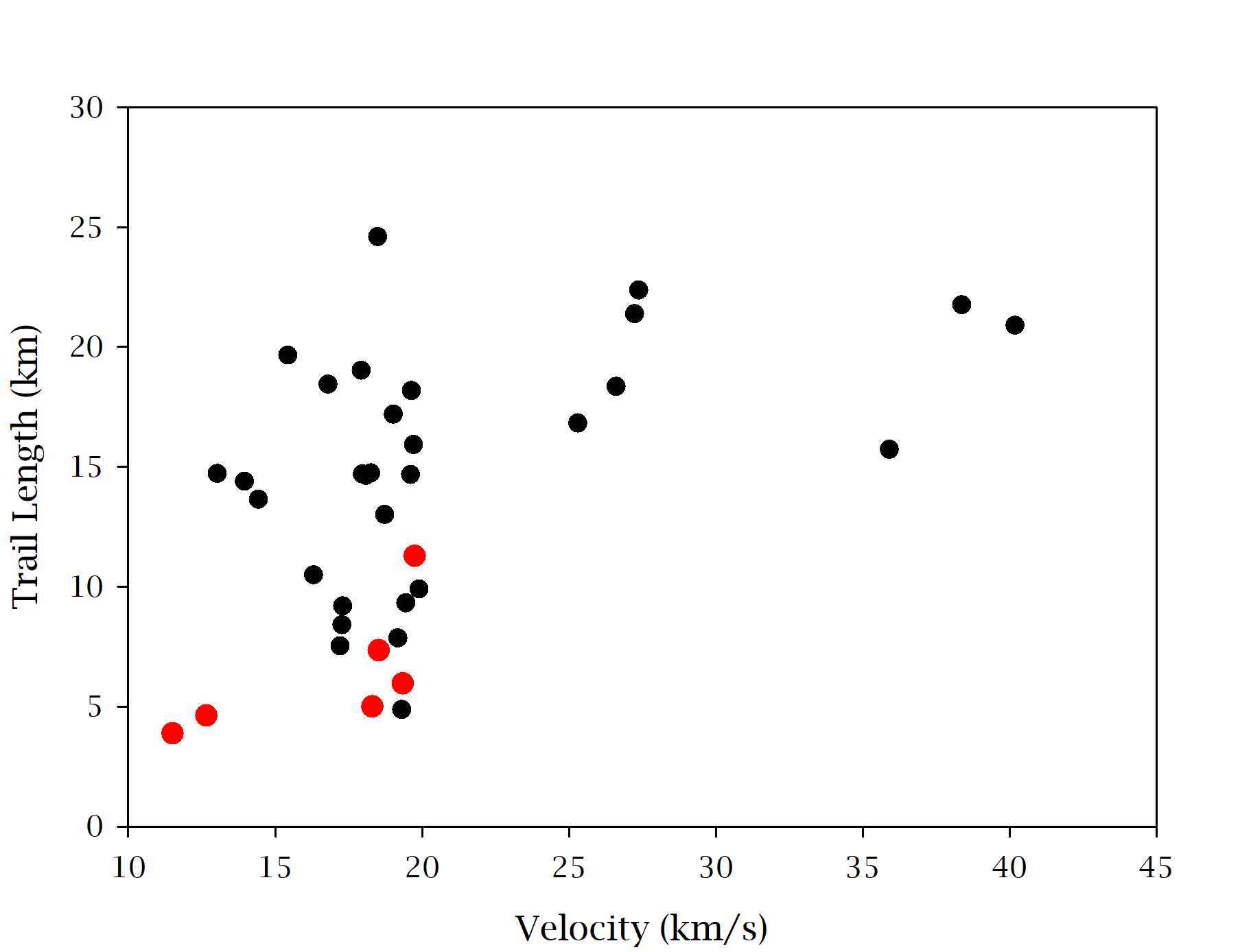}
  \caption{Total meteor trail length as a function of speed. This represents the total length of luminous flight above the detection limit of the instrument (near magnitude +9). Red symbols represent the six events which display sudden onset lightcurves.}
  \label{fig:length-speed}
\end{figure}
\begin{figure}
  \includegraphics[width=\linewidth]{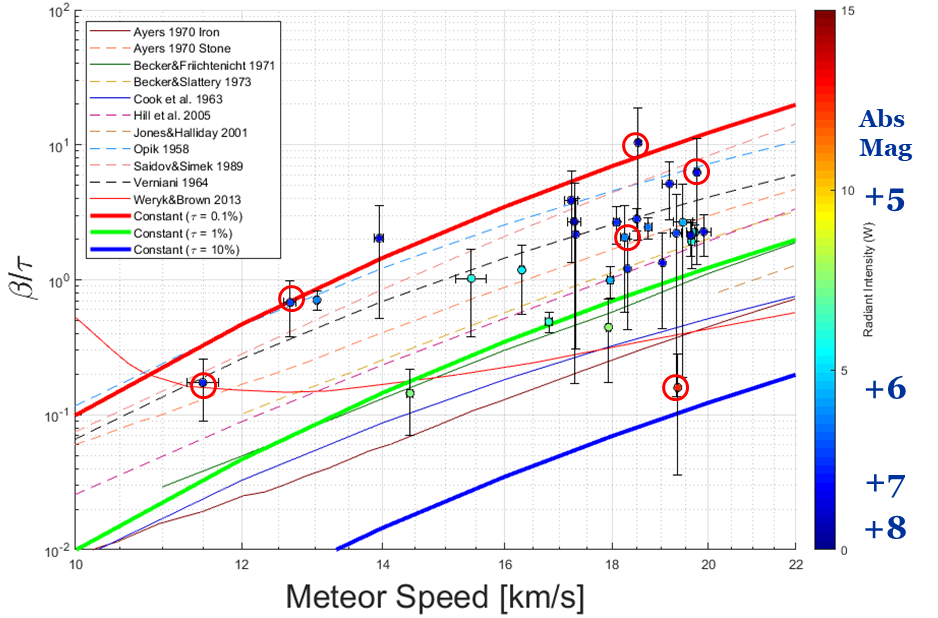}
  \caption{Measured values for $\beta$ / ${\tau}$ for all 28 simultaneous radar-optical events with velocities below 20 km/s as a function of speed. Here $\beta$ is appropriate to a chondritic composition.  Each data point is colour coded according to the radiant intensity at the specular point. The reference curves are the same as in Figure \ref{fig:betatauall}. Circled points (in red) are meteors which have abrupt onset lightcurves, potentially consistent with an iron composition. }
  \label{fig:betatau-iron}
\end{figure}

Figures \ref{fig:ht-speed} and \ref{fig:length-speed} show the distribution of luminous begin heights and trail length as a function of speed for our events. The six sudden onset lightcurve meteors are shown in red. It is clear that these events have systematically lower begin heights than the rest of our population and are among the shortest path lengths. The values for trail length and begin height are in the range adopted by \cite{Capek2019} as representative of iron meteoroids. Figure \ref{fig:betatau-iron} shows $\beta$ / ${\tau}$ for only the lowest speed population (below 20 km/s) with the individual sudden-onset lightcurve events circled. The circled events we interpret as probable iron meteoroids. For such events, the only difference in measured $\beta$ / ${\tau}$ would be that the average mass of ablated atoms should be 56 amu (appropriate to iron) as opposed to our adopted chondritic average of 24 amu.\par
\begin{figure}
  \includegraphics[width=\linewidth]{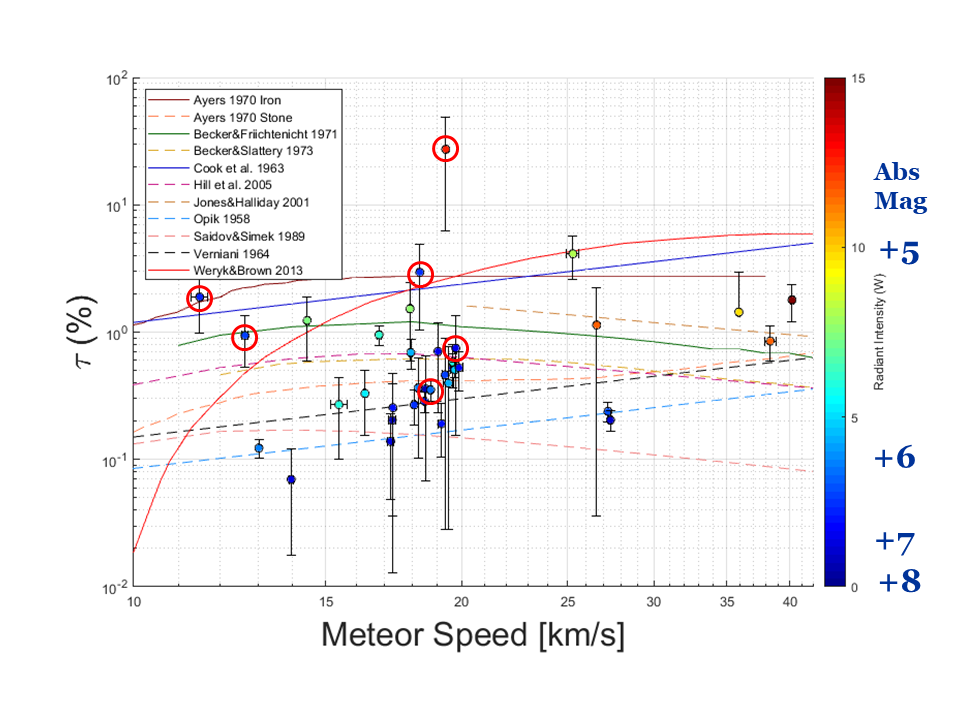}
  \caption{Best estimate for $\tau$ for all events. For circled events, values for $\beta$ appropriate to iron (from \cite{DeLuca2018}) are used. Each meteor is colour-coded by the radiant intensity at the specular point. }
  \label{fig:tau-all}
\end{figure}

Since $\beta$ is larger for pure iron vs. chondritic bodies for a given speed, in Figure ~\ref{fig:tau-all} we show the equivalent ${\tau}$  for all events assuming a pure iron composition for the six circled events and (for those six events) using the $\beta$ relation from \citet{DeLuca2018}. All other events are assumed chondritic. \par
Our values for luminous efficiency show significant scatter, similar to other recent studies (eg. \citet{WerykBrown2013, Subasinghe2018, Capek2019}. We find values of $\tau$ ranging from $<$0.1\% to as high as 30\%. There is no clear speed dependence at our low range of speeds, but there is a slight trend of higher $\tau$ with radiant intensity (see Figure \ref{fig:tauvsI}), a result also found in \cite{WerykBrown2013}, though our number statistics are small so the significance of this trend is questionable. For our slow population (under 20 km/s) the average $\tau$ is 1.5\%, but this is misleading as there is a single larger outlier near 30\% (which is also one of our iron candidates). It is worth noting that for this iron candidate, reprocessing the measurements assuming it to be of chondritic composition drops $\tau$ to under 7\%. Removing this outlier, produces an average $\tau$ near 0.6\%, {with a median value of 0.4\%}, more representative of the slow population. The standard deviation of this group is also 0.6\%. From Figure \ref{fig:tau-all} our six iron meteoroids show $\tau$ values straddling the range given by the lab-based measurements of iron particles reported by \citet{becker1971} .\par
Our events show luminous efficiencies below the values in our prior work \citep{WerykBrown2013}, but we suggest that this may in part be due to the differences in radiant power covered by the two studies as well as the limited speed range in the current work. The earlier study had detection limits roughly one order of magnitude brighter (in radiant power) than the current survey and higher $\tau$ for larger radiant powers are consistent with the trends we find. \par
The variance in $\tau$ almost certainly reflects, in part, the variation in the underlying spectra of the meteors in our sample. Our assumption of a G-band zero magnitude bolometric power of 945W scaled to each meteor ignores this spectral diversity. To probe this relationship in more detail, spectral information, such as that used in the study of \citet{vojacek2019} is required and would be highly desirable for an extension of the present work. {Another source of scatter in $\tau$ may be fragmentation, which can affect the initial radius of meteor trails and ultimately our estimates of the electron line density (and ultimately $\tau$). Currently these estimates use the average initial radius model of \citet{Jones2005a}.}

\par
The accuracy of the electron line densities estimated from the full wave modelling could be improved by performing multi-frequency fits to common echoes. In particular, a subset of slow meteors are detected at all three CMOR frequencies. For such events, model fits varying electron line density, trail radius and the diffusion coefficient to the complete amplitude - time echo records on all three frequencies would provide more stringent constraints and more robust uncertainties. {We plan to examine such multi-frequency common optical events in a future extension of the present work.} \par

\begin{figure}
  \includegraphics[width=\linewidth]{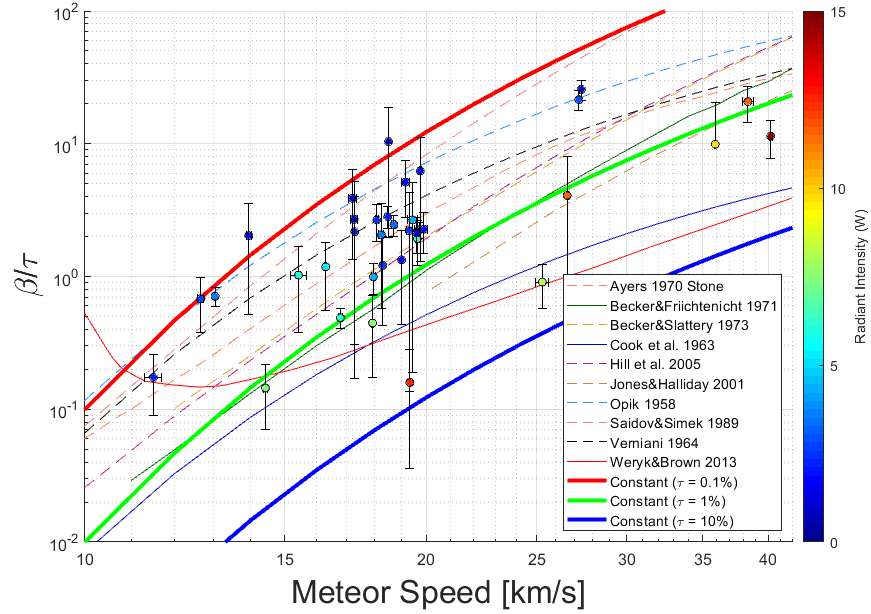}
  \caption{Measured values for $\beta$ / ${\tau}$ for all 36 simultaneous radar-optical events as a function of speed. Each data point is color coded according to the radiant intensity at the specular point, where 10W corresponds approximately to a +5 G-band magnitude meteor. Also shown for comparison are estimates of the ratio of $\beta$ / ${\tau}$ derived from literature reporting various values for $\tau$. Here we use $\beta$ appropriate to a chondritic composition derived from \cite{jones1997} as parameterized in \cite{WerykBrown2013} (their equation 15). Shown are estimates for $\tau$ for stony composition from \citet{Ayers1970}, from laboratory measurements by \citet{becker1971, Becker1973}, from small camera measurements of deceleration and luminosity from \citet{COOK1973, Verniani1964} and from theory by \citet{Opik1955, jones2001}. Also shown are the hybrid theory-lab measurements estimates of $\tau$ from \citet{Hill2005} and direct estimates of the functional velocity form of $\beta$ / ${\tau}$ from optical-radar measurements by \citet{Saidov1989, WerykBrown2013}. Curves of constant $\tau$ are shown for comparison.}
  \label{fig:betatauall}
\end{figure}
	
\begin{sidewaystable}

\small
\centering
\caption{Orbits and trajectory information for all simultaneous optical-radar meteor events. Vel is the average speed of the meteor determined from two station optical records, $q$ is the electron line density at the specular point measured following the approach described in the text, $H_b$ and $H_e$  are the begin and end heights of the optical meteor, while $H_r$ is the radar measured specular height, and Mag is the absolute magnitude of the segment of the trail corresponding to $H_r$. The orbital semi-major axis $a$, eccentricity e, inclination (J2000.0), and perihelion distance q are also listed together with Tisserand's parameter with respect to Jupiter, T$_j$. Entries in bold show abrupt onset lightcurves.}
\label{tab:orbits}
\scalebox{0.8}{
\begin{tabular}{llllllllllll}
\small
Event & $Vel$ & $q$ & $H_b$ & $H_e$ & $H_r$ & $Mag$ & $a$ & $e$ & $i$ & q & $T_j$ \\ 
Name & [km/s] & $10^{13}$[e/m] & [km] & [km] & [km] & [G] & [A.U] &  & [deg] & [A.U] &  \\ \hline\hline
\setrow{\bfseries}20190130:075702 & 11.50 & 1.00 & 80.6 & 78.5 & 79.3 & 6.34 & 0.92 & 0.09 & 5.5 & 0.84 & 6.48 \\ \hline
{\clearrow}
\setrow{\bfseries}20180420:044750 & 12.66 & 3.58 & 85.3 & 83 & 83.3 & 6.37 & 1.13 & 0.2 & 7 & 0.90 & 5.5 \\ \hline
{\clearrow}
20181015:022016 & 13.03 & 4.60 & 86 & 82.6 & 84.7 & 6.07 & 1.73 & 0.43 & 8.78 & 0.99 & 4.04 \\ \hline
20190104:235336 & 13.95 & 4.00 & 91.6 & 82.7 & 90.2 & 6.99 & 1.81 & 0.48 & 6.71 & 0.94 & 3.9 \\ \hline
20180809:032309 & 14.43 & 1.42 & 85.1 & 76.3 & 79.8 & 5.2 & 1.81 & 0.5 & 1.7 & 0.91 & 3.9 \\ \hline
20190104:104246 & 15.43 & 6.70 & 93.2 & 82.7 & 85.9 & 5.5 & 0.77 & 0.41 & 12.68 & 0.45 & 7.45 \\ \hline
20180811:035811 & 16.31 & 5.77 & 94.2 & 88.1 & 91.3 & 5.52 & 1.6 & 0.5 & 0.31 & 0.80 & 4.21 \\ \hline
20170826:042315 & 16.80 & 2.62 & 87.8 & 82.3 & 86.3 & 5.07 & 2.5 & 0.64 & 7.75 & 0.90 & 3.14 \\ \hline
20190404:080544 & 17.21 & 3.70 & 92.8 & 88.6 & 89.9 & 7.21 & 0.68 & 0.58 & 1 & 0.29 & 8.26 \\ \hline
20190508:032428 & 17.27 & 1.30 & 93.7 & 88.1 & 87.8 & 7.68 & 2.22 & 0.61 & 0.98 & 0.87 & 3.37 \\ \hline
20180718:054557 & 17.30 & 3.48 & 92 & 87.2 & 87.3 & 6.09 & 2.07 & 0.59 & 1.4 & 0.85 & 3.54 \\ \hline
20190325:014453 & 17.93 & 2.40 & 94 & 81 & 81.6 & 5.43 & 2.04 & 0.6 & 1.1 & 0.82 & 3.55 \\ \hline
20170827:013652 & 17.96 & 3.02 & 94.9 & 88.9 & 95.2 & 6.49 & 1.78 & 0.57 & 0.96 & 0.77 & 3.88 \\ \hline
20190429:022947 & 18.09 & 4.50 & 96.7 & 89.5 & 91.9 & 6.36 & 1.38 & 0.53 & 2.9 & 0.65 & 4.66 \\ \hline
20190524:040736 & 18.25 & 4.80 & 95.9 & 89.1 & 88.5 & 5.9 & 5.13 & 0.82 & 5.48 & 0.92 & 2.14 \\ \hline
\setrow{\bfseries}20180703:035626 & 18.31 & 1.89 & 87 & 83.6 & 86.3 & 5.4 & 1.23 & 0.46 & 8.9 & 0.66 & 5.07 \\ \hline
{\clearrow}20190325:033020 & 18.49 & 4.06 & 97 & 89.3 & 96.6 & 6.52 & 0.71 & 0.58 & 8.13 & 0.30 & 7.91 \\ \hline
\setrow{\bfseries}20180709:052400 & 18.52 & 3.90 & 87.8 & 82.8 & 82.3 & 7.6 & 1.38 & 0.5 & 6 & 0.69 & 4.63 \\ \hline
{\clearrow}20180716:030941 & 18.73 & 5.00 & 94.1 & 88.9 & 92.4 & 5.99 & 0.84 & 0.55 & 5.1 & 0.38 & 6.86 \\ \hline
20180707:025102 & 19.02 & 1.26 & 96 & 86.2 & 94.3 & 6.15 & 2.48 & 0.68 & 4.5 & 0.79 & 3.11 \\ \hline
20190404:040043 & 19.17 & 6.07 & 92.2 & 87.7 & 89.3 & 6.67 & 0.92 & 0.51 & 12.5 & 0.45 & 6.4 \\ \hline
20180811:071420 & 19.31 & 3.25 & 87.1 & 84.5 & 84.8 & 6.27 & 1.16 & 0.53 & 6.5 & 0.55 & 5.3 \\ \hline
\setrow{\bfseries}20190508:045521 & 19.34 & 1.10 & 87 & 82.8 & 85.3 & 4.58 & 1.03 & 0.47 & 12.1 & 0.55 & 5.82 \\ \hline
{\clearrow}20190521:023120 & 19.45 & 6.40 & 93.6 & 86.7 & 87.6 & 5.94 & 3.58 & 0.76 & 9.7 & 0.86 & 2.51 \\ \hline
20181231:031707 & 19.61 & 1.20 & 96.7 & 90.4 & 97.2 & 7.53 & 0.89 & 0.46 & 18.53 & 0.48 & 6.55 \\ \hline
20180712:054519 & 19.64 & 6.22 & 96 & 88.5 & 93.6 & 5.72 & 0.75 & 0.61 & 7.4 & 0.29 & 7.5 \\ \hline
20180617:025412 & 19.71 & 5.69 & 95 & 90.4 & 92.9 & 5.68 & 1.71 & 0.59 & 5 & 0.70 & 3.96 \\ \hline
\setrow{\bfseries}20180715:035420 & 19.75 & 5.00 & 89.8 & 81.8 & 81.7 & 6.15 & 2.66 & 0.69 & 7.9 & 0.82 & 2.98 \\ \hline
{\clearrow}20180719:042518 & 19.90 & 2.15 & 95.2 & 90.6 & 95.1 & 6.76 & 0.76 & 0.64 & 3.7 & 0.27 & 7.5 \\ \hline
20170702:054951 & 25.30 & 1.82 & 99.3 & 92.3 & 92.8 & 5.71 & 2.73 & 0.79 & 1.2 & 0.57 & 2.79 \\ \hline
20190527:055059 & 26.60 & 10.00 & 101.9 & 90.2 & 92.3 & 4.68 & 2.28 & 0.77 & 12.05 & 0.52 & 3.11 \\ \hline
20190527:025613 & 27.24 & 12.00 & 91.4 & 84.7 & 90.5 & 6.26 & 2.03 & 0.77 & 2.55 & 0.47 & 3.36 \\ \hline
20190526:034610 & 27.38 & 4.90 & 91.8 & 84.9 & 90.0 & 7.44 & 1.73 & 0.76 & 2.34 & 0.42 & 3.75 \\ \hline
20190531:075115 & 35.91 & 8.30 & 103.8 & 96.7 & 98.8 & 5.03 & 2 & 0.91 & 3.63 & 0.18 & 3.12 \\ \hline
20170703:080601 & 38.38 & 17.01 & 95.1 & 86 & 87.8 & 4.08 & 1.65 & 0.95 & 15.4 & 0.08 & 3.5 \\ \hline
20170726:060820 & 40.19 & 16.50 & 100.7 & 91.5 & 96.4 & 3.92 & 1.63 & 0.96 & 30.6 & 0.07 & 3.47 \\ \hline
\end{tabular}}
\end{sidewaystable}

\begin{sidewaystable}
\centering
\small
\caption{Measured values of {\(\frac{\beta}{\tau}\)}. Here the mass is found by integrating the complete lightcurve and applying the tabulated value for $\tau$.}
\label{tab:betatau}
\scalebox{0.8}{
\begin{tabular}{lllllll}
\hline\hline
Event & Vel &{\large${\frac{\beta}{\tau}}$}  & $\pm$ & \large{$\tau$} & $\pm$ & Mass\\ 
Name&[km/s]&&&\%&\%&[kg]\\ \hline\hline
\setrow{\bfseries}20190130:075702 & 11.50 & 0.324 & 0.150 & 1.89 & 0.906 & 5.09E-06 \\ \hline
{\clearrow}
\setrow{\bfseries}20180420:044750 & 12.66 & 1.267 & 0.555 & 0.94 & 0.430 & 1.47E-05 \\ \hline
{\clearrow}20181015:022016 & 13.03 & 0.710 & 0.114 & 0.123 & 0.020 & 3.71E-05 \\ \hline
20190104:235336 & 13.95 & 2.032 & 1.512 & 0.070 & 0.052 & 2.90E-05 \\ \hline
20180809:032309 & 14.43 & 0.144 & 0.075 & 1.235 & 0.638 & 5.70E-06 \\ \hline
20190104:104246 & 15.43 & 1.024 & 0.647 & 0.270 & 0.171 & 1.49E-05 \\ \hline
20180811:035811 & 16.31 & 1.183 & 0.629 & 0.330 & 0.175 & 4.71E-06 \\ \hline
20170826:042315 & 16.80 & 0.490 & 0.085 & 0.951 & 0.164 & 7.35E-06 \\ \hline
20190404:080544 & 17.21 & 3.884 & 2.536 & 0.138 & 0.090 & 1.97E-06 \\ \hline
20190508:032428 & 17.27 & 2.703 & 2.532 & 0.203 & 0.190 & 2.74E-06 \\ \hline
20180718:054557 & 17.30 & 2.173 & 1.865 & 0.255 & 0.219 & 4.24E-06 \\ \hline
20190325:014453 & 17.93 & 0.446 & 0.273 & 1.525 & 0.934 & 4.63E-06 \\ \hline
20170827:013652 & 17.96 & 0.992 & 0.264 & 0.692 & 0.184 & 4.23E-06 \\ \hline
20190429:022947 & 18.09 & 2.667 & 0.820 & 0.268 & 0.082 & 2.99E-06 \\ \hline
20190524:040736 & 18.25 & 2.061 & 1.486 & 0.365 & 0.263 & 4.29E-06 \\ \hline
\setrow{\bfseries}20180703:035626 & 18.31 & 2.259 & 1.462 & 2.96 & 1.92 & 1.39E-06 \\ \hline
{\clearrow}20190325:033020 & 18.49 & 2.819 & 0.525 & 0.286 & 0.053 & 1.35E-05 \\ \hline
\setrow{\bfseries}20180709:052400 & 18.52 & 19.382 & 15.704 & 0.36 & 0.30 & 8.13E-06 \\ \hline
{\clearrow}20180716:030941 & 18.73 & 2.454 & 0.448 & 0.353 & 0.064 & 3.11E-06 \\ \hline
20180707:025102 & 19.02 & 1.336 & 0.898 & 0.705 & 0.474 & 5.89E-06 \\ \hline
20190404:040043 & 19.17 & 5.146 & 2.346 & 0.191 & 0.087 & 4.30E-06 \\ \hline
20180811:071420 & 19.31 & 2.213 & 2.078 & 0.462 & 0.433 & 6.83E-07 \\ \hline
\setrow{\bfseries}20190508:045521 & 19.34 & 0.297 & 0.230 & 27.31 & 21.8 & 2.45E-07 \\ \hline
{\clearrow}20190521:023120 & 19.45 & 2.668 & 2.480 & 0.398 & 0.370 & 3.54E-06 \\ \hline
20181231:031707 & 19.61 & 2.130 & 0.651 & 0.520 & 0.159 & 1.36E-06 \\ \hline
20180712:054519 & 19.64 & 1.927 & 0.722 & 0.579 & 0.217 & 5.23E-06 \\ \hline
20180617:025412 & 19.71 & 2.244 & 0.314 & 0.507 & 0.071 & 2.52E-06 \\ \hline
\setrow{\bfseries}20180715:035420 & 19.75 & 11.65 & 9.23 & 0.75 & 0.60 & 9.55E-06 \\ \hline
{\clearrow}20180719:042518 & 19.90 & 2.273 & 0.784 & 0.526 & 0.181 & 1.20E-06 \\ \hline
20170702:054951 & 25.30 & 0.908 & 0.334 & 4.127 & 1.520 & 2.44E-07 \\ \hline
20190527:055059 & 26.60 & 4.079 & 3.951 & 1.137 & 1.102 & 1.10E-06 \\ \hline
20190527:025613 & 27.24 & 21.457 & 3.717 & 0.239 & 0.041 & 2.08E-06 \\ \hline
20190526:034610 & 27.38 & 25.657 & 4.518 & 0.204 & 0.036 & 1.01E-06 \\ \hline
20190531:075115 & 35.91 & 9.914 & 10.438 & 1.437 & 1.513 & 2.67E-07 \\ \hline
20170703:080601 & 38.38 & 20.783 & 6.349 & 0.850 & 0.260 & 1.18E-06 \\ \hline
20170726:060820 & 40.19 & 11.393 & 3.701 & 1.791 & 0.582 & 7.17E-07 \\ \hline
\end{tabular}}
\end{sidewaystable}

\section{Conclusions}

From a two year study capturing more than one thousand simultaneous optical-radar meteors, 36 events, mostly of low speed, were examined in detail. These meteors had G-band magnitudes of around +6 at the specular point and followed asteroidal orbits. {These  are the first in-situ radar-optical simultaneous measurements at such faint magnitudes and low speeds and therefore represent a critical check on laboratory measurements of both $\beta$  \citep{DeLuca2018} and $\tau$ \citep{Tarnecki2019}.}

Our major conclusions are:
\begin{enumerate}
    \item The ratio of $\beta$ / ${\tau}$ shows a strong {negative } correlation with radiant power. No correlation is found with speed.
    \item Among our slow ($<$ 20 km/s) population, some 20\% showed characteristics consistent with pure iron meteoroids. 
    \item The G-band luminous efficiencies at low speeds show considerable scatter, but average 0.6\% $\pm$ 0.6\%, {with a median value of 0.4\%}.
    \item A slight positive correlation of $\tau$ with speed was identified.
\end{enumerate}

\section{Acknowledgements}
This work was supported in part by the NASA Meteoroid Environment Office under cooperative agreement 80NSSC18M0046. PGB also acknowledges funding support from the Natural Sciences and Engineering Research council of Canada and the Canada Research Chairs program. We thank Z. Krzeminski for help in optical data reduction and J. Gill, M. Mazur for software support and camera operations. 

\bibliography{references}

\begin{thebibliography}{55}
\expandafter\ifx\csname natexlab\endcsname\relax\def\natexlab#1{#1}\fi
\providecommand{\bibinfo}[2]{#2}
\ifx\xfnm\relax \def\xfnm[#1]{\unskip,\space#1}\fi
\bibitem[{Ayers et~al.(1970)Ayers, McCrosky \& Shao}]{Ayers1970}
\bibinfo{author}{Ayers, W.}, \bibinfo{author}{McCrosky, R.}, \&
  \bibinfo{author}{Shao, C. C.-Y.} (\bibinfo{year}{1970}).
\newblock \bibinfo{title}{{Photographic observations of 10 artificial
  meteors}}.
\newblock {\it \bibinfo{journal}{SAO Spec. Rep {\#}317}\/},  {\it
  \bibinfo{volume}{317}\/}, \bibinfo{pages}{40}.
\bibitem[{Becker \& Friichtenicht(1971)}]{becker1971}
\bibinfo{author}{Becker, D.~G.}, \& \bibinfo{author}{Friichtenicht, J.~F.}
  (\bibinfo{year}{1971}).
\newblock \bibinfo{title}{{Measurement and interpretation of the luminous
  efficiencies of iron and copper simulated micrometeors}}.
\newblock {\it \bibinfo{journal}{Astrophysical Journal}\/},  {\it
  \bibinfo{volume}{166}\/}.
\bibitem[{Becker \& Slattery(1973)}]{Becker1973}
\bibinfo{author}{Becker, D.~G.}, \& \bibinfo{author}{Slattery, J.}
  (\bibinfo{year}{1973}).
\newblock \bibinfo{title}{{Luminous efficiency measurements for silicon and
  aluminum simulated micrometeors}}.
\newblock {\it \bibinfo{journal}{Astrophysical Journal}\/},  {\it
  \bibinfo{volume}{186}\/}, \bibinfo{pages}{1127--1139}.
\bibitem[{Borovi{\v{c}}ka(1993)}]{Borovicka1993}
\bibinfo{author}{Borovi{\v{c}}ka, J.} (\bibinfo{year}{1993}).
\newblock \bibinfo{title}{{A fireball spectrum analysis}}.
\newblock {\it \bibinfo{journal}{Astronomy and Astrophysics}\/},  {\it
  \bibinfo{volume}{279}\/}, \bibinfo{pages}{627--645}.
\bibitem[{Borovi{\v{c}}ka(1994)}]{Borovicka1994a}
\bibinfo{author}{Borovi{\v{c}}ka, J.} (\bibinfo{year}{1994}).
\newblock \bibinfo{title}{{Two components in meteor spectra}}.
\newblock {\it \bibinfo{journal}{Planetary and Space Science}\/},  {\it
  \bibinfo{volume}{42}\/}, \bibinfo{pages}{145--150}.
\bibitem[{Borovi{\v{c}}ka et~al.(2005)Borovi{\v{c}}ka, Koten, Spurn{\'{y}},
  Bocek \& Stork}]{Borovicka2005}
\bibinfo{author}{Borovi{\v{c}}ka, J.}, \bibinfo{author}{Koten, P.},
  \bibinfo{author}{Spurn{\'{y}}, P.}, \bibinfo{author}{Bocek, J.}, \&
  \bibinfo{author}{Stork, R.} (\bibinfo{year}{2005}).
\newblock \bibinfo{title}{{A survey of meteor spectra and orbits: evidence for
  three populations of Na-free meteoroids}}.
\newblock {\it \bibinfo{journal}{Icarus}\/},  {\it \bibinfo{volume}{174}\/},
  \bibinfo{pages}{15--30}.
\bibitem[{Brown et~al.(2017)Brown, Stober, Schult, Krzeminski, Cooke \&
  Chau}]{Brown2017}
\bibinfo{author}{Brown, P.~G.}, \bibinfo{author}{Stober, G.},
  \bibinfo{author}{Schult, C.}, \bibinfo{author}{Krzeminski, Z.},
  \bibinfo{author}{Cooke, W.}, \& \bibinfo{author}{Chau, J.}
  (\bibinfo{year}{2017}).
\newblock \bibinfo{title}{{Simultaneous optical and meteor head echo
  measurements using the Middle Atmosphere Alomar Radar System (MAARSY): Data
  collection and preliminary analysis}}.
\newblock {\it \bibinfo{journal}{Planetary and Space Science}\/},  {\it
  \bibinfo{volume}{141}\/}, \bibinfo{pages}{25--34}.
\bibitem[{Brown et~al.(2008)Brown, Weryk, Wong \& Jones}]{Brown2008}
\bibinfo{author}{Brown, P.~G.}, \bibinfo{author}{Weryk, R.},
  \bibinfo{author}{Wong, D.}, \& \bibinfo{author}{Jones, J.}
  (\bibinfo{year}{2008}).
\newblock \bibinfo{title}{{A meteoroid stream survey using the Canadian Meteor
  Orbit Radar I. Methodology and radiant catalogue}}.
\newblock {\it \bibinfo{journal}{Icarus}\/},  {\it \bibinfo{volume}{195}\/},
  \bibinfo{pages}{317--339}.
\bibitem[{Campbell-Brown(2015)}]{Campbell-Brown2015}
\bibinfo{author}{Campbell-Brown, M.~D.} (\bibinfo{year}{2015}).
\newblock \bibinfo{title}{{A population of small refractory meteoroids in
  asteroidal orbits}}.
\newblock {\it \bibinfo{journal}{Planetary and Space Science}\/},  {\it
  \bibinfo{volume}{118}\/}, \bibinfo{pages}{8--13}.
\bibitem[{Campbell-Brown et~al.(2012)Campbell-Brown, Kero, Szasz,
  Pellinen-Wannberg \& Weryk}]{Campbell-Brown2012}
\bibinfo{author}{Campbell-Brown, M.~D.}, \bibinfo{author}{Kero, J.},
  \bibinfo{author}{Szasz, C.}, \bibinfo{author}{Pellinen-Wannberg, a.}, \&
  \bibinfo{author}{Weryk, R.} (\bibinfo{year}{2012}).
\newblock \bibinfo{title}{{Photometric and ionization masses of meteors with
  simultaneous EISCAT UHF radar and intensified video observations}}.
\newblock {\it \bibinfo{journal}{Journal of Geophysical Research}\/},  {\it
  \bibinfo{volume}{117}\/}, \bibinfo{pages}{1--13}.
\bibitem[{{\v{C}}apek \& Borovi{\v{c}}ka(2017)}]{capekboro2017}
\bibinfo{author}{{\v{C}}apek, D.}, \& \bibinfo{author}{Borovi{\v{c}}ka, J.}
  (\bibinfo{year}{2017}).
\newblock \bibinfo{title}{{Ablation of small iron meteoroids–First results}}.
\newblock {\it \bibinfo{journal}{Planetary and Space Science}\/},  {\it
  \bibinfo{volume}{143}\/}, \bibinfo{pages}{159--163}.
\bibitem[{Capek et~al.(2019)Capek, Koten, Borovi{\v{c}}ka, Voj{\'{a}}{\v{c}}ek,
  Spurn{\'{y}} \& {\v{S}}tork}]{Capek2019}
\bibinfo{author}{Capek, D.}, \bibinfo{author}{Koten, P.},
  \bibinfo{author}{Borovi{\v{c}}ka, J.}, \bibinfo{author}{Voj{\'{a}}{\v{c}}ek,
  V.}, \bibinfo{author}{Spurn{\'{y}}, P.}, \& \bibinfo{author}{{\v{S}}tork, R.}
  (\bibinfo{year}{2019}).
\newblock \bibinfo{title}{{Small iron meteoroids}}.
\newblock {\it \bibinfo{journal}{Astronomy {\&} Astrophysics}\/},  {\it
  \bibinfo{volume}{625}\/}, \bibinfo{pages}{A106}.
\bibitem[{Carrillo-S{\'{a}}nchez et~al.(2016)Carrillo-S{\'{a}}nchez,
  Nesvorn{\'{y}}, Pokorn{\'{y}}, Janches \& Plane}]{Carrillo-Sanchez2016}
\bibinfo{author}{Carrillo-S{\'{a}}nchez, J.~D.},
  \bibinfo{author}{Nesvorn{\'{y}}, D.}, \bibinfo{author}{Pokorn{\'{y}}, P.},
  \bibinfo{author}{Janches, D.}, \& \bibinfo{author}{Plane, J.}
  (\bibinfo{year}{2016}).
\newblock \bibinfo{title}{{Sources of cosmic dust in the Earth's atmosphere}}.
\newblock {\it \bibinfo{journal}{Geophysical Research Letters}\/},  {\it
  \bibinfo{volume}{43}\/}, \bibinfo{pages}{979--11}.
\bibitem[{Ceplecha et~al.(1998)Ceplecha, Spalding, Jacobs, ReVelle, Tagliaferri
  \& Brown}]{Ceplecha1998}
\bibinfo{author}{Ceplecha, Z.}, \bibinfo{author}{Spalding, R.~E.},
  \bibinfo{author}{Jacobs, C.}, \bibinfo{author}{ReVelle, D.},
  \bibinfo{author}{Tagliaferri, E.}, \& \bibinfo{author}{Brown, P.~G.}
  (\bibinfo{year}{1998}).
\newblock \bibinfo{title}{{Superbolides}}.
\newblock {\it \bibinfo{journal}{Meteoroids 1998}\/},  (p.
  \bibinfo{pages}{37–54}).
\bibitem[{Close et~al.(2004)Close, Oppenheim, Hunt \& Coster}]{Close2004}
\bibinfo{author}{Close, S.}, \bibinfo{author}{Oppenheim, M.~M.},
  \bibinfo{author}{Hunt, S.}, \& \bibinfo{author}{Coster, A.}
  (\bibinfo{year}{2004}).
\newblock \bibinfo{title}{{A technique for calculating meteor plasma density
  and meteoroid mass from radar head echo scattering}}.
\newblock {\it \bibinfo{journal}{Icarus}\/},  {\it \bibinfo{volume}{168}\/},
  \bibinfo{pages}{43--52}.
\bibitem[{Cook et~al.(1973)Cook, FORTI, McCrosky, Posen, Southworth \&
  WILLIAMS}]{COOK1973}
\bibinfo{author}{Cook, A.}, \bibinfo{author}{FORTI, G.},
  \bibinfo{author}{McCrosky, R.}, \bibinfo{author}{Posen, A.},
  \bibinfo{author}{Southworth, R.~B.}, \& \bibinfo{author}{WILLIAMS, J.}
  (\bibinfo{year}{1973}).
\newblock \bibinfo{title}{{Combined observations of meteors by image-orthicon
  television camera and multi-station radar(to compare ionization with
  luminosity)}}.
\newblock In {\it \bibinfo{booktitle}{NASA, Washington Evolutionary and Phys.
  Properties of Meteoroids p 23-44(SEE N 74-19436 10-30)}\/}.
\bibitem[{DeLuca et~al.(2018)DeLuca, Munsat, Thomas \& Sternovsky}]{DeLuca2018}
\bibinfo{author}{DeLuca, M.}, \bibinfo{author}{Munsat, T.},
  \bibinfo{author}{Thomas, E.}, \& \bibinfo{author}{Sternovsky, Z.}
  (\bibinfo{year}{2018}).
\newblock \bibinfo{title}{{The ionization efficiency of aluminum and iron at
  meteoric velocities}}.
\newblock {\it \bibinfo{journal}{Planetary and Space Science}\/},  {\it
  \bibinfo{volume}{156}\/}, \bibinfo{pages}{111--116}.
\bibitem[{Friichtenicht et~al.(1968)Friichtenicht, Slattery \&
  {E}}]{Friichtenicht1968}
\bibinfo{author}{Friichtenicht, J.~F.}, \bibinfo{author}{Slattery, J.}, \&
  \bibinfo{author}{{E}} (\bibinfo{year}{1968}).
\newblock \bibinfo{title}{{A laboratory measurement of meteor luminous
  efficiency}}.
\newblock {\it \bibinfo{journal}{The Astrophysical}\/},  {\it
  \bibinfo{volume}{151}\/}, \bibinfo{pages}{747--758}.
\bibitem[{Gural(2016)}]{gural2016}
\bibinfo{author}{Gural, P.} (\bibinfo{year}{2016}).
\newblock \bibinfo{title}{{A fast meteor detection algorithm}}.
\newblock In {\it \bibinfo{booktitle}{International Meteor Conference Egmond,
  the Netherlands, 2-5 June 2016}\/} (pp. \bibinfo{pages}{96--104}).
\bibitem[{Hill et~al.(2005)Hill, Rogers \& Hawkes}]{Hill2005}
\bibinfo{author}{Hill, K.}, \bibinfo{author}{Rogers, L.}, \&
  \bibinfo{author}{Hawkes, R.} (\bibinfo{year}{2005}).
\newblock \bibinfo{title}{{High geocentric velocity meteor ablation}}.
\newblock {\it \bibinfo{journal}{Astronomy and Astrophysics}\/},  {\it
  \bibinfo{volume}{444}\/}, \bibinfo{pages}{615–624}.
\bibitem[{Jarosewich(1990)}]{jarosewich1990}
\bibinfo{author}{Jarosewich, E.} (\bibinfo{year}{1990}).
\newblock \bibinfo{title}{{Jarosewich (1990) - Chemical analyses of meteorites-
  A compilation of stony and iron meteorite analyses}}.
\newblock {\it \bibinfo{journal}{Meteoritics}\/},  {\it
  \bibinfo{volume}{25}\/}, \bibinfo{pages}{323--337}.
\bibitem[{Jones et~al.(2005)Jones, Brown, Ellis, Webster, Campbell-Brown,
  Krzemenski \& Weryk}]{Jones2005}
\bibinfo{author}{Jones, J.}, \bibinfo{author}{Brown, P.~G.},
  \bibinfo{author}{Ellis, K.~J.}, \bibinfo{author}{Webster, A.},
  \bibinfo{author}{Campbell-Brown, M.~D.}, \bibinfo{author}{Krzemenski, Z.}, \&
  \bibinfo{author}{Weryk, R.} (\bibinfo{year}{2005}).
\newblock \bibinfo{title}{{The Canadian Meteor Orbit Radar : system overview
  and preliminary results}}.
\newblock {\it \bibinfo{journal}{Planetary and Space Science}\/},  {\it
  \bibinfo{volume}{53}\/}, \bibinfo{pages}{413--421}.
\bibitem[{Jones \& Campbell-Brown(2005)}]{Jones2005a}
\bibinfo{author}{Jones, J.}, \& \bibinfo{author}{Campbell-Brown, M.~D.}
  (\bibinfo{year}{2005}).
\newblock \bibinfo{title}{{The initial train radius of sporadic meteors}}.
\newblock {\it \bibinfo{journal}{Mon. Not. R. Astron. Soc}\/},  {\it
  \bibinfo{volume}{359}\/}, \bibinfo{pages}{1131–1136}.
\bibitem[{Jones et~al.(1998)Jones, Webster \& Hocking}]{jones1998}
\bibinfo{author}{Jones, J.}, \bibinfo{author}{Webster, A.}, \&
  \bibinfo{author}{Hocking, W.} (\bibinfo{year}{1998}).
\newblock \bibinfo{title}{{An improved interferometer design for use with
  meteor radars}}.
\newblock {\it \bibinfo{journal}{Radio Science}\/},  {\it
  \bibinfo{volume}{33}\/}, \bibinfo{pages}{55–65}.
\bibitem[{Jones(1995)}]{jones1995}
\bibinfo{author}{Jones, W.} (\bibinfo{year}{1995}).
\newblock \bibinfo{title}{{Theory of the initial radius of meteor trains}}.
\newblock {\it \bibinfo{journal}{Monthly Notices of the Royal Astronomical
  Society}\/},  {\it \bibinfo{volume}{275}\/}, \bibinfo{pages}{812--818}.
\bibitem[{Jones(1997)}]{jones1997}
\bibinfo{author}{Jones, W.} (\bibinfo{year}{1997}).
\newblock \bibinfo{title}{{Theoretical and observational determinations of the
  ionization coefficient of meteors}}.
\newblock {\it \bibinfo{journal}{Monthly Notices of the Royal Astronomical
  Society}\/},  {\it \bibinfo{volume}{288}\/}, \bibinfo{pages}{995–1003}.
\bibitem[{Jones \& Halliday(2001)}]{jones2001}
\bibinfo{author}{Jones, W.}, \& \bibinfo{author}{Halliday, I.}
  (\bibinfo{year}{2001}).
\newblock \bibinfo{title}{{Effects of excitation and ionization in meteor
  trains}}.
\newblock {\it \bibinfo{journal}{Monthly Notices of the Royal Astronomical
  Society}\/},  {\it \bibinfo{volume}{320}\/}, \bibinfo{pages}{417--423}.
\bibitem[{Jordi et~al.(2010)Jordi, Gebran, Carrasco, de~Bruijne, Voss,
  Fabricius, Knude, Vallenari, Kohley \& Mora}]{Jordi2010}
\bibinfo{author}{Jordi, C.}, \bibinfo{author}{Gebran, M.},
  \bibinfo{author}{Carrasco, J.~M.}, \bibinfo{author}{de~Bruijne, J.},
  \bibinfo{author}{Voss, H.}, \bibinfo{author}{Fabricius, C.},
  \bibinfo{author}{Knude, J.}, \bibinfo{author}{Vallenari, A.},
  \bibinfo{author}{Kohley, R.}, \& \bibinfo{author}{Mora, A.}
  (\bibinfo{year}{2010}).
\newblock \bibinfo{title}{{Gaia broad band photometry}}.
\newblock {\it \bibinfo{journal}{Astronomy {\&} Astrophysics}\/},  {\it
  \bibinfo{volume}{523}\/}, \bibinfo{pages}{A48}.
\bibitem[{Kaiser \& Closs(1952)}]{Kaiser1952}
\bibinfo{author}{Kaiser, T.}, \& \bibinfo{author}{Closs, R.~L.}
  (\bibinfo{year}{1952}).
\newblock \bibinfo{title}{{Theory of radio reflections from meteor trails: I}}.
\newblock {\it \bibinfo{journal}{Phil Mag}\/},  {\it \bibinfo{volume}{43}\/},
  \bibinfo{pages}{1--32}.
\bibitem[{Klekociuk et~al.(2005)Klekociuk, Brown, Pack, ReVelle, Edwards,
  Spalding, Tagliaferri, Yoo \& Zagari}]{Klekociuk2005}
\bibinfo{author}{Klekociuk, A.~R.}, \bibinfo{author}{Brown, P.~G.},
  \bibinfo{author}{Pack, D.}, \bibinfo{author}{ReVelle, D.},
  \bibinfo{author}{Edwards, W.}, \bibinfo{author}{Spalding, R.~E.},
  \bibinfo{author}{Tagliaferri, E.}, \bibinfo{author}{Yoo, B.~B.}, \&
  \bibinfo{author}{Zagari, J.} (\bibinfo{year}{2005}).
\newblock \bibinfo{title}{{Meteoritic dust from the atmospheric disintegration
  of a large meteoroid.}}
\newblock {\it \bibinfo{journal}{Nature}\/},  {\it \bibinfo{volume}{436}\/},
  \bibinfo{pages}{1132--5}.
\bibitem[{Liou et~al.(1995)Liou, Dermott \& Xu}]{Liou1995b}
\bibinfo{author}{Liou, J.}, \bibinfo{author}{Dermott, S.}, \&
  \bibinfo{author}{Xu, Y.} (\bibinfo{year}{1995}).
\newblock \bibinfo{title}{{The contribution of cometary dust to the zodiacal
  cloud}}.
\newblock {\it \bibinfo{journal}{Planetary and Space Science}\/},  {\it
  \bibinfo{volume}{43}\/}, \bibinfo{pages}{717--722}.
\bibitem[{Marshall et~al.(2017)Marshall, Brown \& Close}]{Marshall2017}
\bibinfo{author}{Marshall, R.~A.}, \bibinfo{author}{Brown, P.~G.}, \&
  \bibinfo{author}{Close, S.} (\bibinfo{year}{2017}).
\newblock \bibinfo{title}{{Plasma distributions in meteor head echoes and
  implications for radar cross section interpretation}}.
\newblock {\it \bibinfo{journal}{Planetary and Space Science}\/},  (pp.
  \bibinfo{pages}{1--6}).
\bibitem[{Mathews et~al.(2010)Mathews, Briczinski, Malhotra \&
  Cross}]{Mathews2010}
\bibinfo{author}{Mathews, J.}, \bibinfo{author}{Briczinski, S.},
  \bibinfo{author}{Malhotra, A.}, \& \bibinfo{author}{Cross, J.}
  (\bibinfo{year}{2010}).
\newblock \bibinfo{title}{{Extensive meteoroid fragmentation in V/UHF radar
  meteor observations at Arecibo Observatory}}.
\newblock {\it \bibinfo{journal}{Geophysical Research Letters}\/},  {\it
  \bibinfo{volume}{37}\/}, \bibinfo{pages}{1--5}.
\bibitem[{Nesvorn{\'{y}} et~al.(2010)Nesvorn{\'{y}}, Jenniskens, Levison,
  Bottke, Vokrouhlick{\'{y}} \& Gounelle}]{Nesvorny2010}
\bibinfo{author}{Nesvorn{\'{y}}, D.}, \bibinfo{author}{Jenniskens, P.},
  \bibinfo{author}{Levison, H.}, \bibinfo{author}{Bottke, W.},
  \bibinfo{author}{Vokrouhlick{\'{y}}, D.}, \& \bibinfo{author}{Gounelle, M.}
  (\bibinfo{year}{2010}).
\newblock \bibinfo{title}{{Cometary Origin of the Zodiacal Cloud and
  Carbonaceous Micrometeorites. Implications for Hot Debris Disks}}.
\newblock {\it \bibinfo{journal}{The Astrophysical Journal}\/},  {\it
  \bibinfo{volume}{713}\/}, \bibinfo{pages}{816--836}.
\bibitem[{Nishimura et~al.(2001)Nishimura, Sato, Nakamura \&
  Ueda}]{Nishimura2001}
\bibinfo{author}{Nishimura, K.}, \bibinfo{author}{Sato, T.},
  \bibinfo{author}{Nakamura, T.}, \& \bibinfo{author}{Ueda, M.}
  (\bibinfo{year}{2001}).
\newblock \bibinfo{title}{{High Sensitivity Radar-Optical Observations of Faint
  Meteors}}.
\newblock {\it \bibinfo{journal}{IEICE Trans Commun}\/},  {\it
  \bibinfo{volume}{E84-C}\/}, \bibinfo{pages}{1877--1884}.
\bibitem[{Opik(1955)}]{Opik1955}
\bibinfo{author}{Opik, E.~J.} (\bibinfo{year}{1955}).
\newblock \bibinfo{title}{{Meteor radiation, ionization and atomic luminous
  efficiency}}.
\newblock {\it \bibinfo{journal}{Proceedings of the Royal Society of London.
  Series A, Mathematical and Physical Sciences}\/},  {\it
  \bibinfo{volume}{230}\/}, \bibinfo{pages}{463–501}.
\bibitem[{Poulter \& Baggaley(1977)}]{Poulter1977}
\bibinfo{author}{Poulter, E.}, \& \bibinfo{author}{Baggaley, W.}
  (\bibinfo{year}{1977}).
\newblock \bibinfo{title}{{Radiowave scattering from meteoric ionization}}.
\newblock {\it \bibinfo{journal}{Journal of Atmospheric and Terrestrial
  Physics}\/},  {\it \bibinfo{volume}{39}\/}, \bibinfo{pages}{757--768}.
\bibitem[{Saidov \& Simek(1989)}]{Saidov1989}
\bibinfo{author}{Saidov, K.}, \& \bibinfo{author}{Simek, M.}
  (\bibinfo{year}{1989}).
\newblock \bibinfo{title}{{Luminous efficiency coefficient from simultaneous
  meteor observations}}.
\newblock {\it \bibinfo{journal}{Bulletin of the Astronomical Institutes of
  Czechoslovakia}\/},  {\it \bibinfo{volume}{40}\/},
  \bibinfo{pages}{330–332}.
\bibitem[{Silber et~al.(2015)Silber, Brown \& Krzeminski}]{Silber2015}
\bibinfo{author}{Silber, E.}, \bibinfo{author}{Brown, P.~G.}, \&
  \bibinfo{author}{Krzeminski, Z.} (\bibinfo{year}{2015}).
\newblock \bibinfo{title}{{Optical observations of meteors generating
  infrasound: Weak shock theory and validation}}.
\newblock {\it \bibinfo{journal}{Journal of Geophysical Research: Planets}\/},
  (pp. \bibinfo{pages}{413--428}).
\bibitem[{Slattery \& Friichtenicht(1967)}]{Slattery1967}
\bibinfo{author}{Slattery, J.}, \& \bibinfo{author}{Friichtenicht, J.~F.}
  (\bibinfo{year}{1967}).
\newblock \bibinfo{title}{{Ionization probability of iron particles at meteoric
  velocities}}.
\newblock {\it \bibinfo{journal}{The Astrophysical Journal}\/},  {\it
  \bibinfo{volume}{147}\/}.
\bibitem[{Subasinghe \& Campbell-Brown(2018)}]{Subasinghe2018}
\bibinfo{author}{Subasinghe, D.}, \& \bibinfo{author}{Campbell-Brown, M.}
  (\bibinfo{year}{2018}).
\newblock \bibinfo{title}{{Luminous Efficiency Estimates of Meteors. II.
  Application to Canadian Automated Meteor Observatory Meteor Events}}.
\newblock {\it \bibinfo{journal}{The Astronomical Journal}\/},  {\it
  \bibinfo{volume}{155}\/}, \bibinfo{pages}{88}.
\bibitem[{Subasinghe et~al.(2016)Subasinghe, Campbell-Brown \&
  Stokan}]{Subasinghe2016}
\bibinfo{author}{Subasinghe, D.}, \bibinfo{author}{Campbell-Brown, M.~D.}, \&
  \bibinfo{author}{Stokan, E.} (\bibinfo{year}{2016}).
\newblock \bibinfo{title}{{Physical characteristics of faint meteors by light
  curve and high-resolution observations, and the implications for parent
  bodies}}.
\newblock {\it \bibinfo{journal}{Monthly Notices of the Royal Astronomical
  Society}\/},  {\it \bibinfo{volume}{457}\/}, \bibinfo{pages}{1289--1298}.
\bibitem[{Subasinghe et~al.(2017)Subasinghe, Campbell-Brown \&
  Stokan}]{Subasinghe2017}
\bibinfo{author}{Subasinghe, D.}, \bibinfo{author}{Campbell-Brown, M.~D.}, \&
  \bibinfo{author}{Stokan, E.} (\bibinfo{year}{2017}).
\newblock \bibinfo{title}{{Luminous efficiency estimates of meteors -I.
  Uncertainty analysis}}.
\newblock {\it \bibinfo{journal}{Planetary and Space Science}\/},  {\it
  \bibinfo{volume}{143}\/}, \bibinfo{pages}{71--77}.
\bibitem[{Tarnecki et~al.(2019)Tarnecki, Marshall, Sternovsky, Munsat \&
  DeLuca}]{Tarnecki2019}
\bibinfo{author}{Tarnecki, L.~K.}, \bibinfo{author}{Marshall, R.~A.},
  \bibinfo{author}{Sternovsky, Z.}, \bibinfo{author}{Munsat, T.~L.}, \&
  \bibinfo{author}{DeLuca, M.} (\bibinfo{year}{2019}).
\newblock \bibinfo{title}{{Laboratory Dust Ablation Experiments to Characterize
  Meteoric Luminous Efficiencies}}.
\newblock In {\it \bibinfo{booktitle}{AGUFM}\/} (pp.
  \bibinfo{pages}{P21F--3436}).
\newblock volume \bibinfo{volume}{2019}.
\bibitem[{Thomas et~al.(2016)Thomas, Hor{\'{a}}nyi, Janches, Munsat, Simolka \&
  Sternovsky}]{Thomas2016}
\bibinfo{author}{Thomas, E.}, \bibinfo{author}{Hor{\'{a}}nyi, M.},
  \bibinfo{author}{Janches, D.}, \bibinfo{author}{Munsat, T.},
  \bibinfo{author}{Simolka, J.}, \& \bibinfo{author}{Sternovsky, Z.}
  (\bibinfo{year}{2016}).
\newblock \bibinfo{title}{{Measurements of the ionization coefficient of
  simulated iron micrometeoroids}}.
\newblock {\it \bibinfo{journal}{Geophysical Research Letters}\/},  {\it
  \bibinfo{volume}{43}\/}, \bibinfo{pages}{3645--3652}.
\bibitem[{Verniani \& Hawkins(1964)}]{Verniani1964}
\bibinfo{author}{Verniani, F.}, \& \bibinfo{author}{Hawkins, G.}
  (\bibinfo{year}{1964}).
\newblock \bibinfo{title}{{On the ionizaing efficiency of meteors}}.
\newblock {\it \bibinfo{journal}{Astrophysical Journal}\/},  {\it
  \bibinfo{volume}{140}\/}, \bibinfo{pages}{1590}.
\bibitem[{Vida et~al.(2019)Vida, Gural, Brown, Campbell-Brown \&
  Wiegert}]{Vida2019a}
\bibinfo{author}{Vida, D.}, \bibinfo{author}{Gural, P.~S.},
  \bibinfo{author}{Brown, P.~G.}, \bibinfo{author}{Campbell-Brown, M.}, \&
  \bibinfo{author}{Wiegert, P.} (\bibinfo{year}{2019}).
\newblock \bibinfo{title}{{Estimating trajectories of meteors: an observational
  Monte Carlo approach - I. Theory}}.
\newblock {\it \bibinfo{journal}{Monthly Notices of the Royal Astronomical
  Society}\/},  {\it \bibinfo{volume}{19}\/}, \bibinfo{pages}{1--19}.
\bibitem[{Voj{\'{a}}{\v{c}}ek et~al.(2019)Voj{\'{a}}{\v{c}}ek, Borovi{\v{c}}ka,
  Koten, Spurn{\'{y}} \& {\v{S}}tork}]{vojacek2019}
\bibinfo{author}{Voj{\'{a}}{\v{c}}ek, V.}, \bibinfo{author}{Borovi{\v{c}}ka,
  J.}, \bibinfo{author}{Koten, P.}, \bibinfo{author}{Spurn{\'{y}}, P.}, \&
  \bibinfo{author}{{\v{S}}tork, R.} (\bibinfo{year}{2019}).
\newblock \bibinfo{title}{{Properties of small meteoroids studied by meteor
  video observations}}.
\newblock {\it \bibinfo{journal}{Astronomy and Astrophysics}\/},  {\it
  \bibinfo{volume}{621}\/}, \bibinfo{pages}{1--21}.
\bibitem[{Webster et~al.(2004)Webster, Brown, Jones, Ellis \&
  Campbell-Brown}]{Webster2004}
\bibinfo{author}{Webster, A.}, \bibinfo{author}{Brown, P.~G.},
  \bibinfo{author}{Jones, J.}, \bibinfo{author}{Ellis, K.~J.}, \&
  \bibinfo{author}{Campbell-Brown, M.~D.} (\bibinfo{year}{2004}).
\newblock \bibinfo{title}{{Canadian Meteor Orbit Radar (CMOR)}}.
\newblock {\it \bibinfo{journal}{Atmospheric Chemistry and Physics}\/},  {\it
  \bibinfo{volume}{4}\/}, \bibinfo{pages}{679–684}.
\bibitem[{Weryk \& Brown(2012)}]{werykbrown2012}
\bibinfo{author}{Weryk, R.}, \& \bibinfo{author}{Brown, P.~G.}
  (\bibinfo{year}{2012}).
\newblock \bibinfo{title}{{Simultaneous radar and video meteors—I: Metric
  comparisons}}.
\newblock {\it \bibinfo{journal}{Planetary and Space Science}\/},  {\it
  \bibinfo{volume}{62}\/}, \bibinfo{pages}{132--152}.
\bibitem[{Weryk \& Brown(2013{\natexlab{a}})}]{Weryk2013a}
\bibinfo{author}{Weryk, R.}, \& \bibinfo{author}{Brown, P.~G.}
  (\bibinfo{year}{2013}{\natexlab{a}}).
\newblock \bibinfo{title}{{Simultaneous radar and video meteors - II:
  Photometry and ionisation}}.
\newblock {\it \bibinfo{journal}{Planetary and Space Science}\/},  {\it
  \bibinfo{volume}{81}\/}, \bibinfo{pages}{32--47}.
\bibitem[{Weryk et~al.(2007)Weryk, Brown, Domokos, Edwards, Krzeminski, Nudds
  \& Welch}]{Weryk2007}
\bibinfo{author}{Weryk, R.}, \bibinfo{author}{Brown, P.~G.},
  \bibinfo{author}{Domokos, A.}, \bibinfo{author}{Edwards, W.},
  \bibinfo{author}{Krzeminski, Z.}, \bibinfo{author}{Nudds, S.~H.}, \&
  \bibinfo{author}{Welch, D.~L.} (\bibinfo{year}{2007}).
\newblock \bibinfo{title}{{The Southern Ontario All-sky Meteor Camera
  Network}}.
\newblock {\it \bibinfo{journal}{Earth, Moon, and Planets}\/},  {\it
  \bibinfo{volume}{102}\/}, \bibinfo{pages}{241--246}.
\bibitem[{Weryk et~al.(2013)Weryk, Campbell-Brown, Wiegert, Brown, Krzeminski
  \& Musci}]{Weryk2013}
\bibinfo{author}{Weryk, R.}, \bibinfo{author}{Campbell-Brown, M.~D.},
  \bibinfo{author}{Wiegert, P.}, \bibinfo{author}{Brown, P.~G.},
  \bibinfo{author}{Krzeminski, Z.}, \& \bibinfo{author}{Musci, R.}
  (\bibinfo{year}{2013}).
\newblock \bibinfo{title}{{The Canadian Automated Meteor Observatory (CAMO):
  System overview}}.
\newblock {\it \bibinfo{journal}{Icarus}\/},  {\it \bibinfo{volume}{225}\/},
  \bibinfo{pages}{614--622}.
\bibitem[{Weryk \& Brown(2013{\natexlab{b}})}]{WerykBrown2013}
\bibinfo{author}{Weryk, R.~J.}, \& \bibinfo{author}{Brown, P.~G.}
  (\bibinfo{year}{2013}{\natexlab{b}}).
\newblock \bibinfo{title}{{Simultaneous radar and video meteors—II:
  Photometry and ionisation}}.
\newblock {\it \bibinfo{journal}{Planetary and Space Science}\/},  {\it
  \bibinfo{volume}{81}\/}, \bibinfo{pages}{32--47}.
\bibitem[{Yang \& Ishiguro(2015)}]{Yang2015}
\bibinfo{author}{Yang, H.}, \& \bibinfo{author}{Ishiguro, M.}
  (\bibinfo{year}{2015}).
\newblock \bibinfo{title}{{Origin of Interplanetary Dust Through Optical
  Properties of Zodiacal Light}}.
\newblock {\it \bibinfo{journal}{The Astrophysical Journal}\/},  {\it
  \bibinfo{volume}{813}\/}, \bibinfo{pages}{87}.

\end{thebibliography}

\newpage

\appendix

\section{Lightcurves}\label{appendix:lightcurves}
In the following pages are plots of the absolute brightness of each meteor as a function of the EMCCD camera time from both stations for common events. In some cases more than two cameras detected the same event; in those cases all three (or four) lightcurves are shown. Note that the radar site is located at Tavistock. The name convention for each event is the time of appearance as YYYYMMDD:HHMMSS in UTC. 

\begin{figure}
     \centering
     \captionsetup[subfigure]{labelformat=empty}
     \begin{subfigure}[b]{0.48\textwidth}
         \centering
         \includegraphics[width=\textwidth]{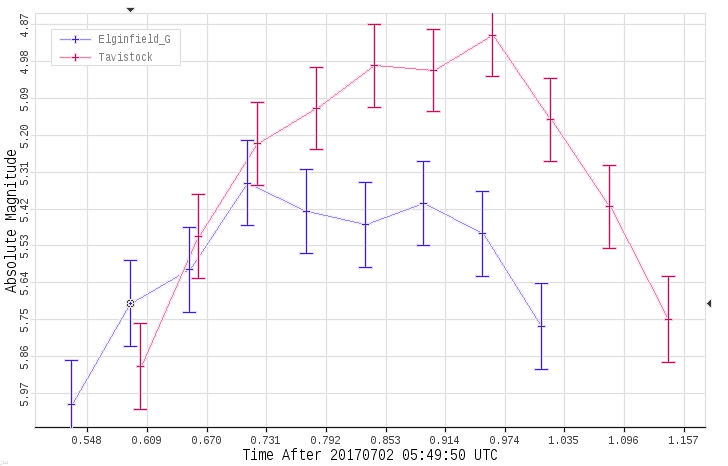}
         \caption{20170702:054951}
         \label{fig:20170702:054951}
     \end{subfigure}
     \begin{subfigure}[b]{0.48\textwidth}
         \centering
         \includegraphics[width=\textwidth]{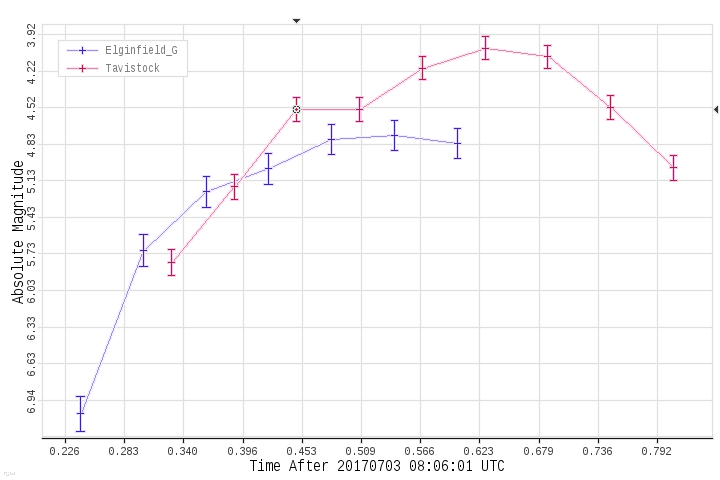}
         \caption{20170703:080601}
         \label{fig:20170703:080601}
           \end{subfigure}
           \medskip
    \begin{subfigure}[b]{0.48\textwidth}
         \centering
         \includegraphics[width=\textwidth]{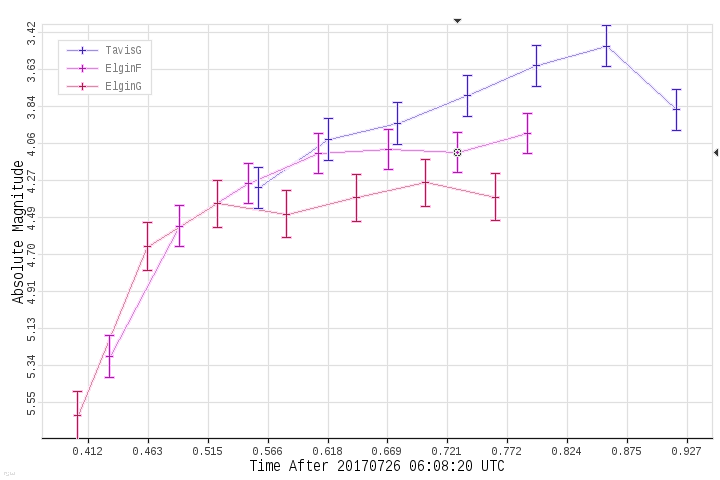}
         \caption{20170726:060820}
         \label{fig:20170702:054951}
     \end{subfigure}
     \begin{subfigure}[b]{0.48\textwidth}
         \centering
         \includegraphics[width=\textwidth]{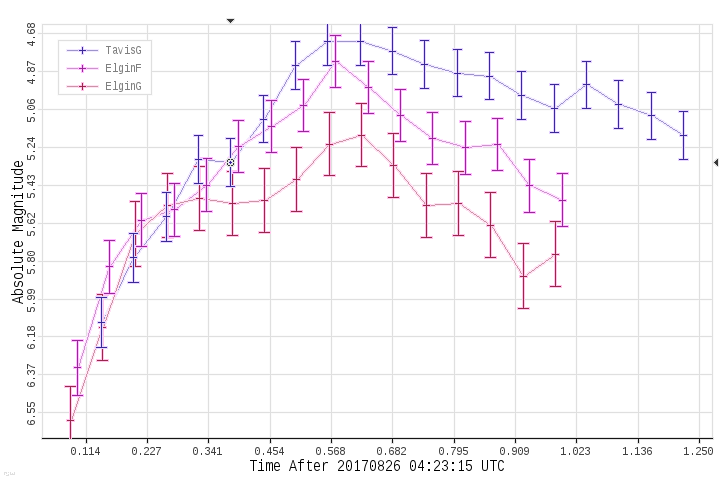}
         \caption{20170826:04231}
         \label{fig:20170703:080601}
     \end{subfigure}
     \begin{subfigure}[b]{0.48\textwidth}
         \centering
         \includegraphics[width=\textwidth]{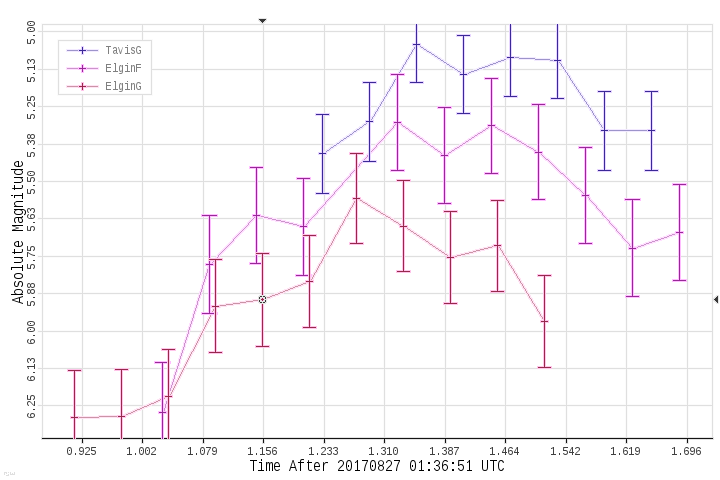}
         \caption{20170827:013651}
         \label{fig:20170702:054951}
     \end{subfigure}
     \begin{subfigure}[b]{0.48\textwidth}
         \centering
         \includegraphics[width=\textwidth]{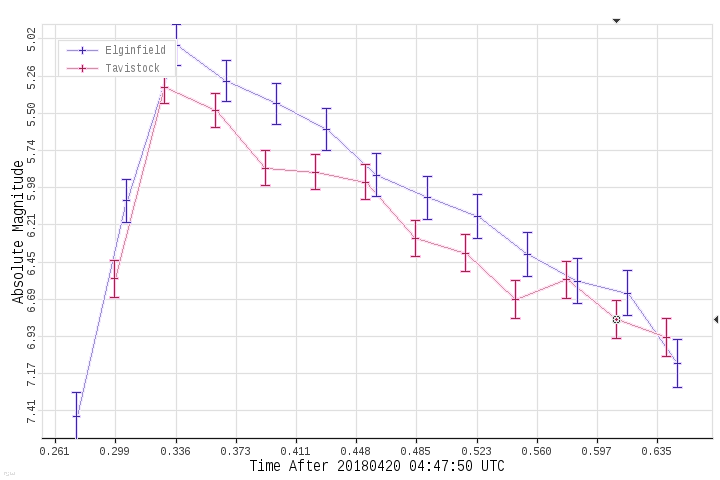}
         \caption{20180420:044750}
         \label{fig:20170703:080601}
         \end{subfigure}
     \end{figure}
 \clearpage       
\begin{figure}[t!]
     \centering
     \captionsetup[subfigure]{labelformat=empty}
    \begin{subfigure}[b]{0.48\textwidth}
         \centering
         \includegraphics[width=\textwidth]{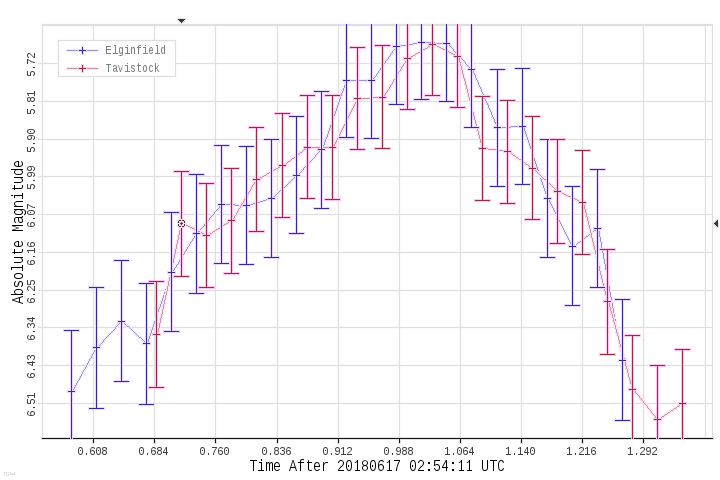}
         \caption{20180617:025411}
         \label{fig:20170702:054951}
     \end{subfigure}
     \begin{subfigure}[b]{0.48\textwidth}
         \centering
         \includegraphics[width=\textwidth]{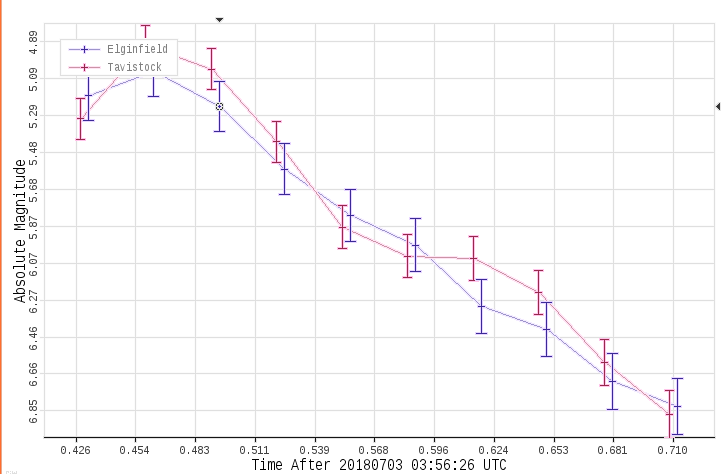}
         \caption{20180703:035626}
         \label{fig:20170703:080601}
     \end{subfigure}
         \begin{subfigure}[b]{0.48\textwidth}
         \centering
         \includegraphics[width=\textwidth]{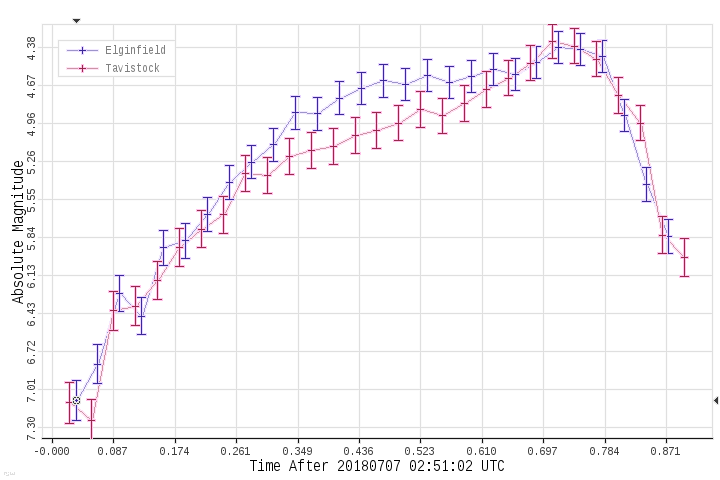}
         \caption{20180617:025411}
         \label{fig:20170702:054951}
     \end{subfigure}
     \begin{subfigure}[b]{0.48\textwidth}
         \centering
         \includegraphics[width=\textwidth]{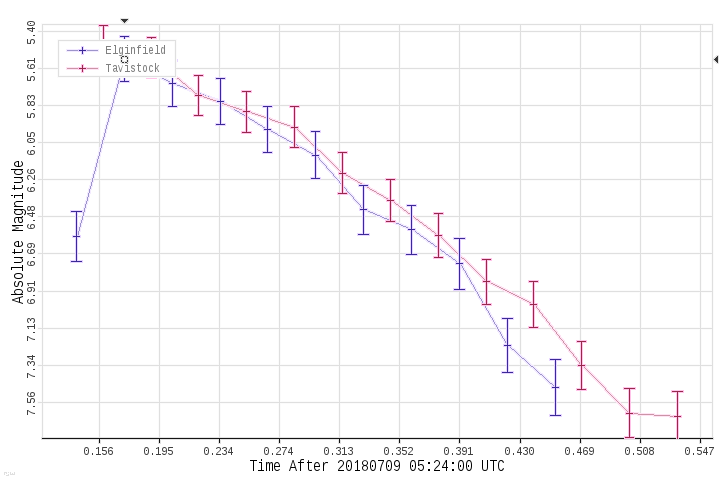}
         \caption{20180709:052400}
         \label{fig:20170703:080601}
     \end{subfigure}
          \begin{subfigure}[b]{0.48\textwidth}
         \centering
         \includegraphics[width=\textwidth]{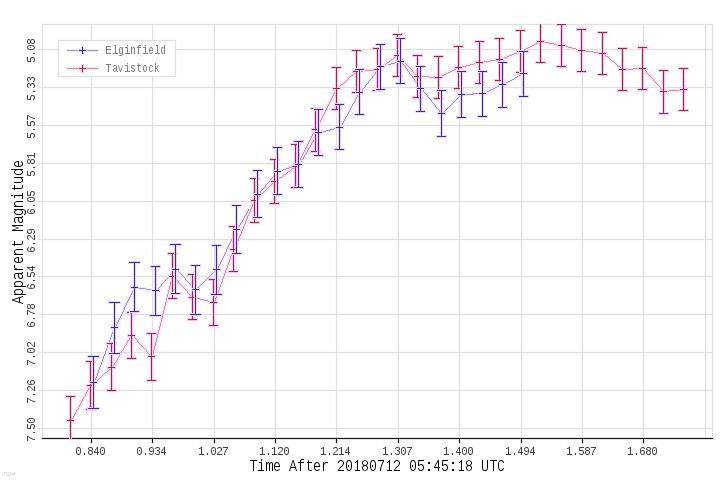}
         \caption{20180712:054518}
         \label{fig:20170702:054951}
     \end{subfigure}
     \begin{subfigure}[b]{0.48\textwidth}
         \centering
         \includegraphics[width=\textwidth]{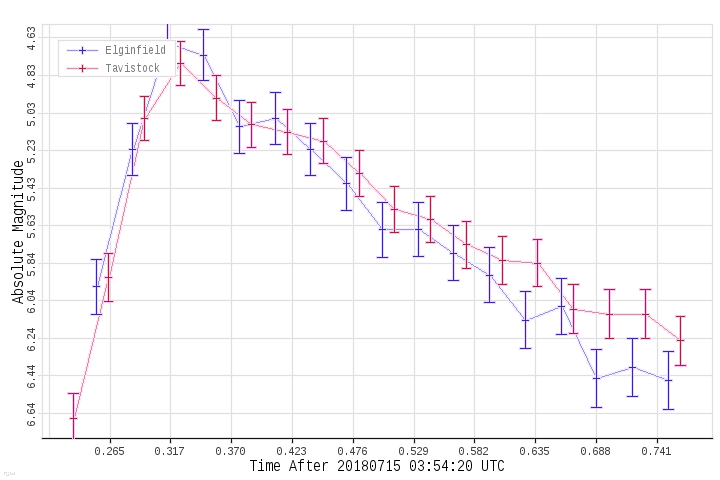}
         \caption{20180715:035420}
         \label{fig:20170703:080601}
     \end{subfigure}
     
 \end{figure}
  \clearpage  
\begin{figure}[t!]
     \centering
     \captionsetup[subfigure]{labelformat=empty}
    \begin{subfigure}[b]{0.48\textwidth}
         \centering
         \includegraphics[width=\textwidth]{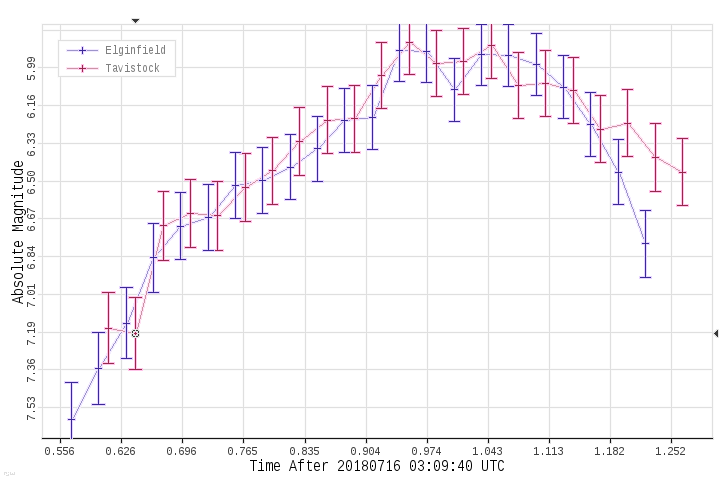}
         \caption{20180716:030940}
         \label{fig:20170702:054951}
     \end{subfigure}
     \begin{subfigure}[b]{0.48\textwidth}
         \centering
         \includegraphics[width=\textwidth]{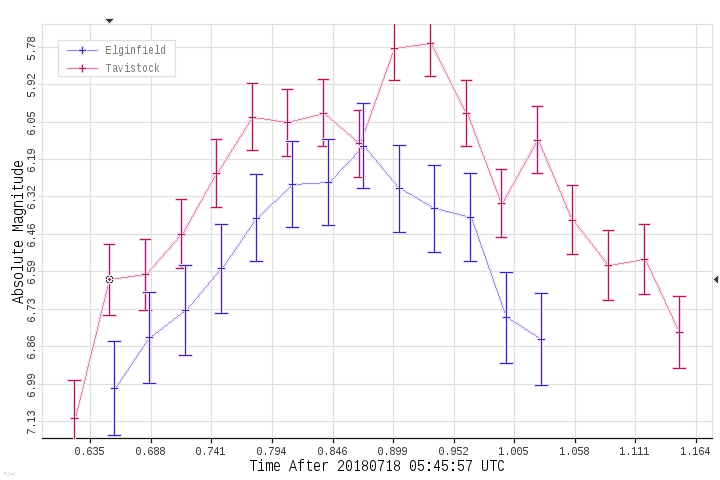}
         \caption{20180718:054557}
         \label{fig:20170703:080601}
     \end{subfigure}
         \begin{subfigure}[b]{0.48\textwidth}
         \centering
         \includegraphics[width=\textwidth]{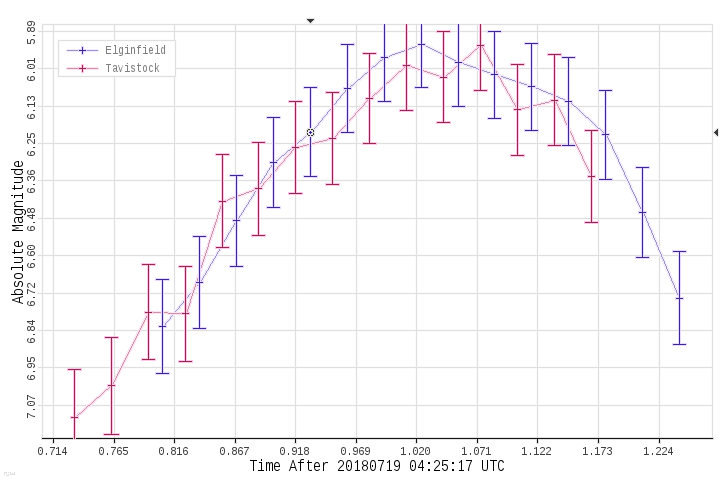}
         \caption{20180719:042517}
         \label{fig:20170702:054951}
     \end{subfigure}
     \begin{subfigure}[b]{0.48\textwidth}
         \centering
         \includegraphics[width=\textwidth]{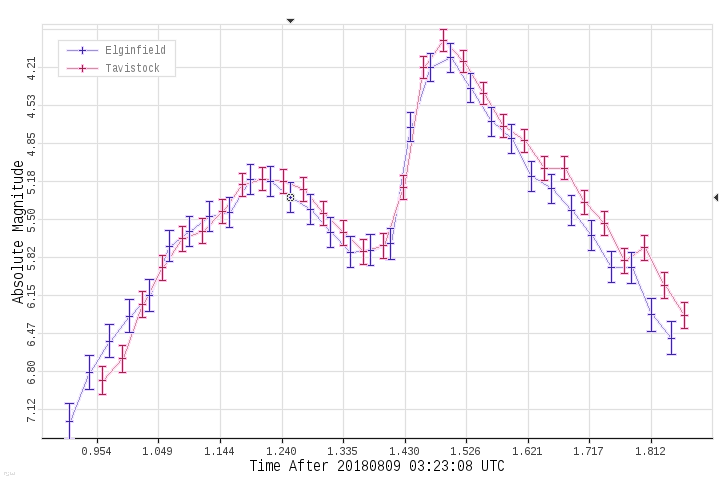}
         \caption{20180809:032308}
         \label{fig:20170703:080601}
     \end{subfigure}
          \begin{subfigure}[b]{0.48\textwidth}
         \centering
         \includegraphics[width=\textwidth]{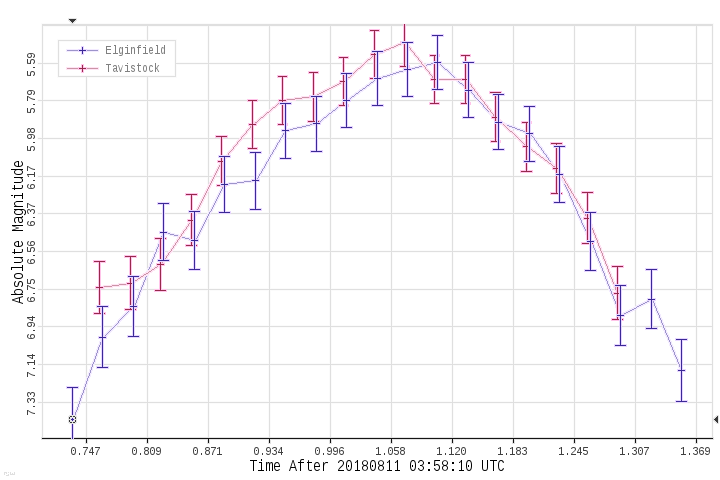}
         \caption{20180811:035810}
         \label{fig:20170702:054951}
     \end{subfigure}
     \begin{subfigure}[b]{0.48\textwidth}
         \centering
         \includegraphics[width=\textwidth]{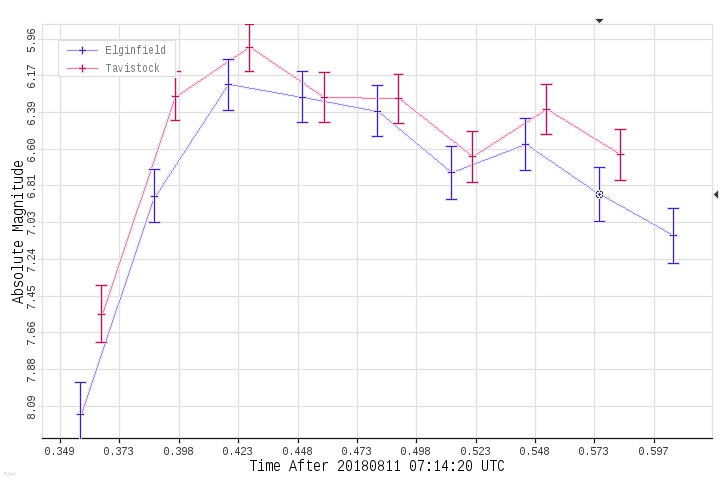}
         \caption{20180811:071420}
         \label{fig:20170703:080601}
     \end{subfigure}
     \end{figure}
       \clearpage  
       \begin{figure}[t!]
     \centering
     \captionsetup[subfigure]{labelformat=empty}
    \begin{subfigure}[b]{0.48\textwidth}
         \centering
         \includegraphics[width=\textwidth]{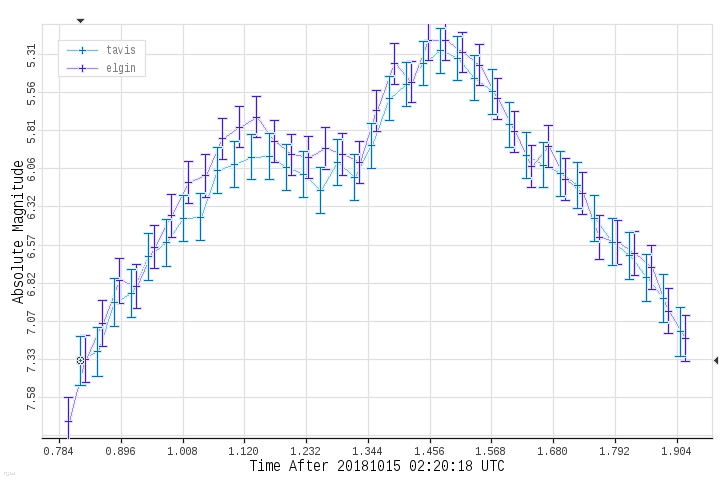}
         \caption{20181015:022018}
         \label{fig:20170702:054951}
     \end{subfigure}
     \begin{subfigure}[b]{0.48\textwidth}
         \centering
         \includegraphics[width=\textwidth]{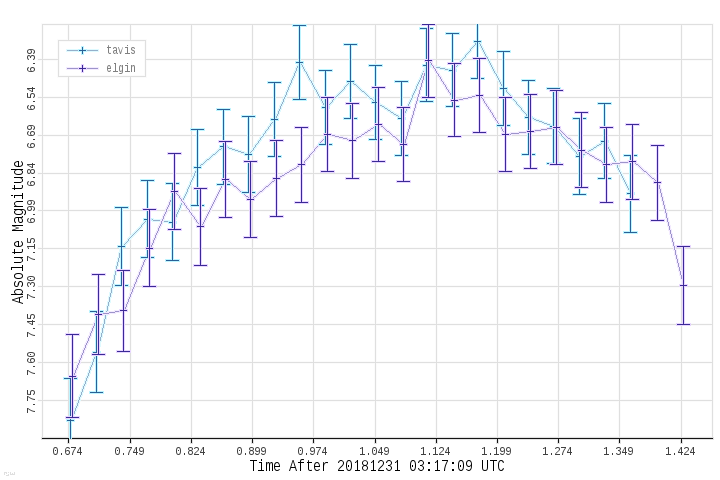}
         \caption{20181231:031709}
         \label{fig:20170703:080601}
     \end{subfigure}
         \begin{subfigure}[b]{0.48\textwidth}
         \centering
         \includegraphics[width=\textwidth]{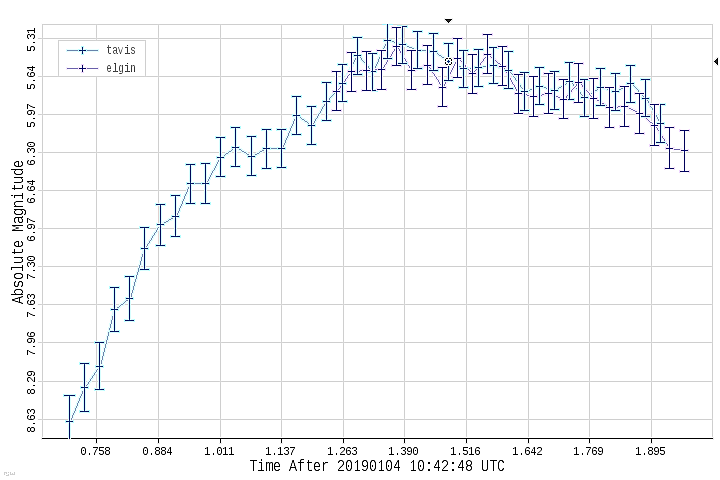}
         \caption{20190104:104248}
         \label{fig:20170702:054951}
     \end{subfigure}
     \begin{subfigure}[b]{0.48\textwidth}
         \centering
         \includegraphics[width=\textwidth]{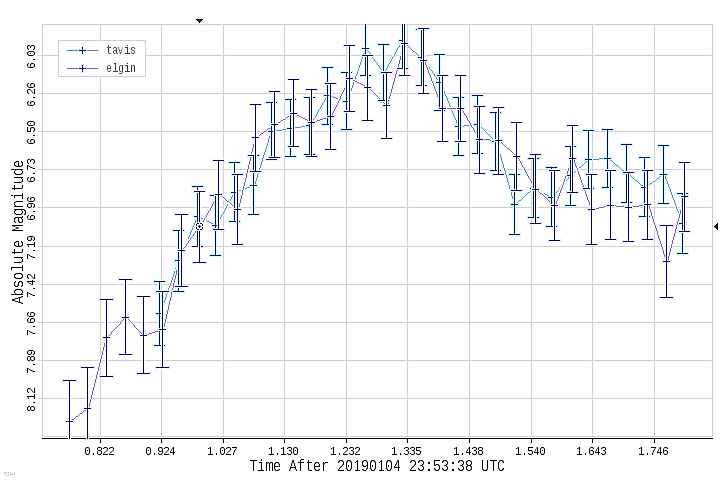}
         \caption{20190104:235338}
         \label{fig:20170703:080601}
     \end{subfigure}
          \begin{subfigure}[b]{0.48\textwidth}
         \centering
         \includegraphics[width=\textwidth]{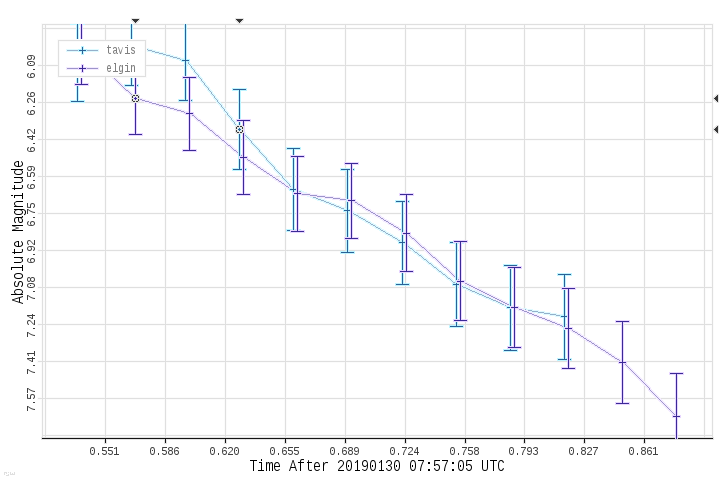}
         \caption{20190130:07570}
         \label{fig:20170702:054951}
     \end{subfigure}
     \begin{subfigure}[b]{0.48\textwidth}
         \centering
         \includegraphics[width=\textwidth]{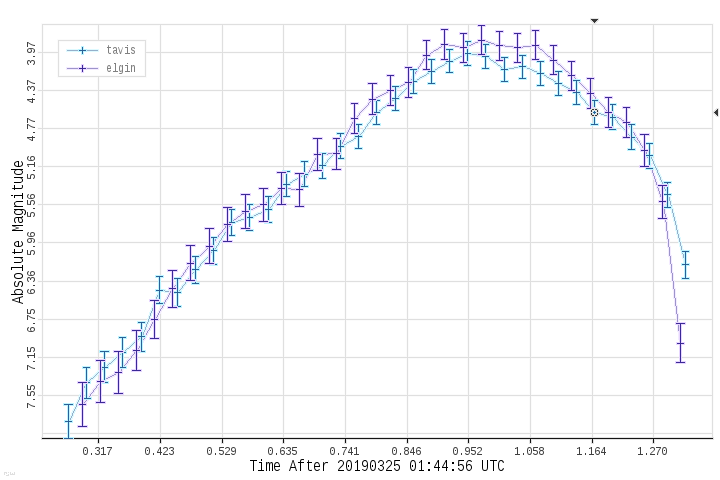}
         \caption{20190325:014456}
         \label{fig:20170703:080601}
     \end{subfigure}
     \end{figure}
     
      \begin{figure}[t!]
     \centering
     \captionsetup[subfigure]{labelformat=empty}
    \begin{subfigure}[b]{0.48\textwidth}
         \centering
         \includegraphics[width=\textwidth]{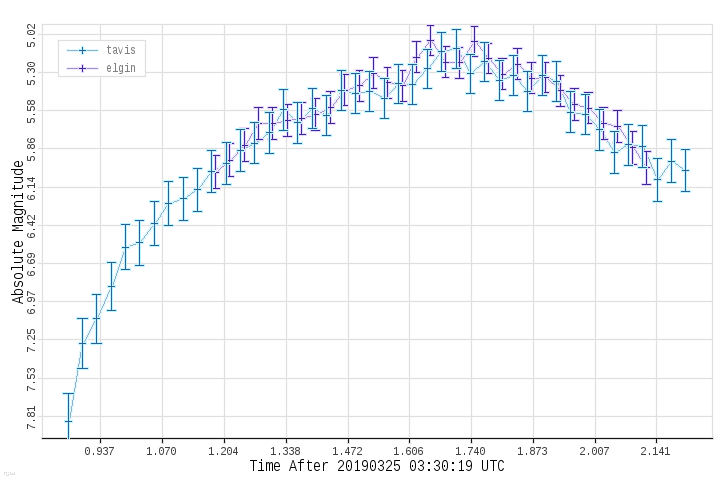}
         \caption{20190325:033019}
         \label{fig:20170702:054951}
     \end{subfigure}
     \begin{subfigure}[b]{0.48\textwidth}
         \centering
         \includegraphics[width=\textwidth]{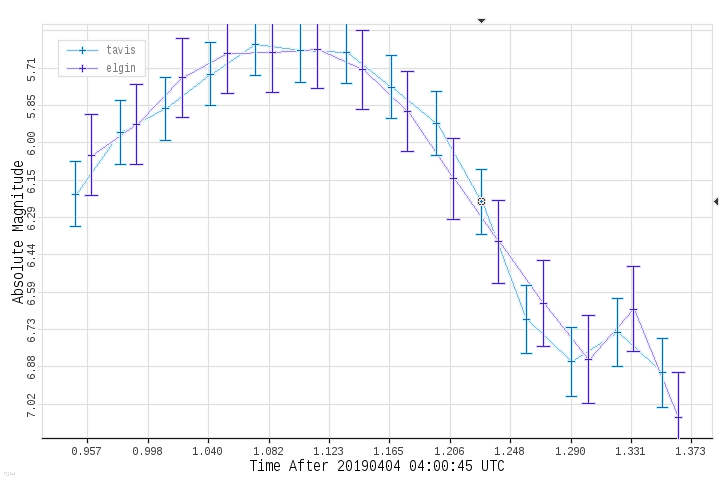}
         \caption{20190404:040045}
         \label{fig:20170703:080601}
     \end{subfigure}
         \begin{subfigure}[b]{0.48\textwidth}
         \centering
         \includegraphics[width=\textwidth]{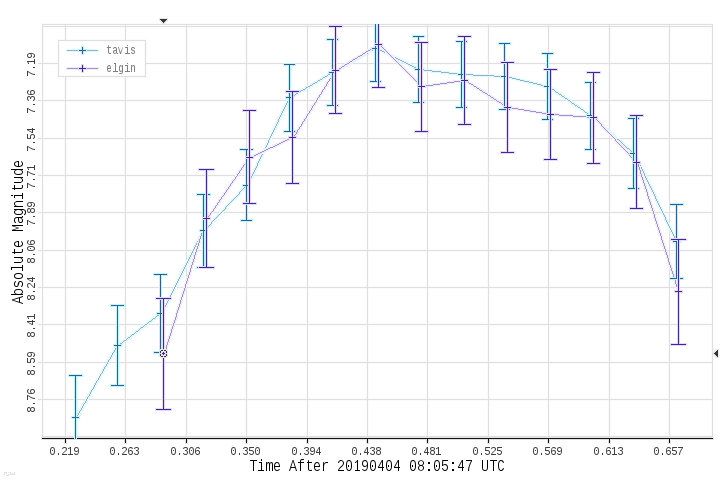}
         \caption{20190404:080547}
         \label{fig:20170702:054951}
     \end{subfigure}
     \begin{subfigure}[b]{0.48\textwidth}
         \centering
         \includegraphics[width=\textwidth]{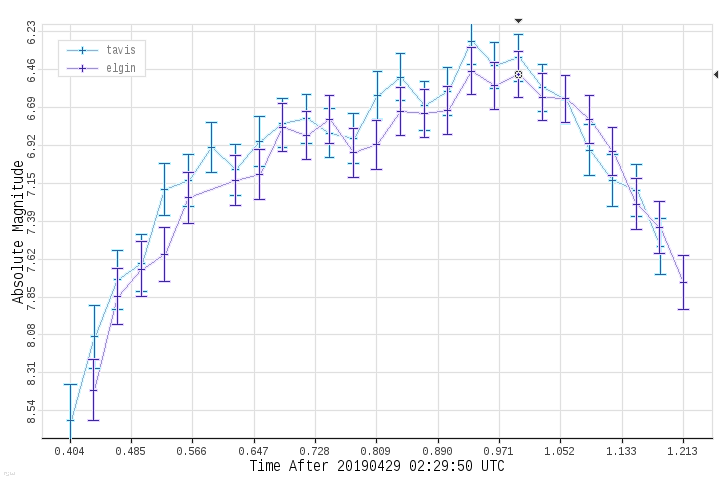}
         \caption{20190429:022950}
         \label{fig:20170703:080601}
     \end{subfigure}
          \begin{subfigure}[b]{0.48\textwidth}
         \centering
         \includegraphics[width=\textwidth]{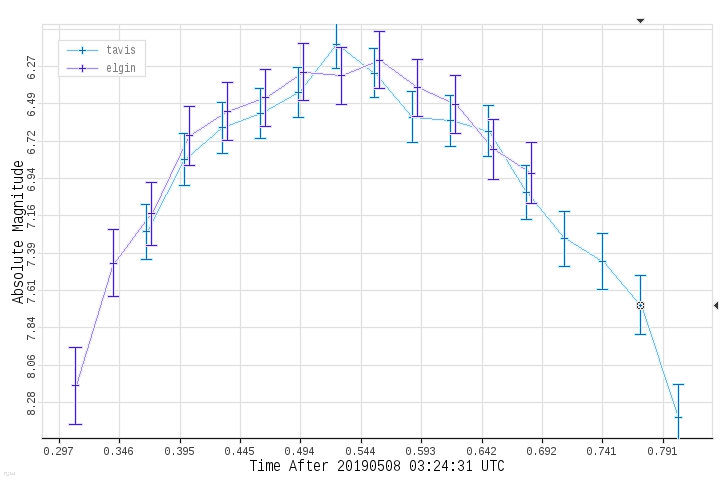}
         \caption{20190508:032431}
         \label{fig:20170702:054951}
     \end{subfigure}
     \begin{subfigure}[b]{0.48\textwidth}
         \centering
         \includegraphics[width=\textwidth]{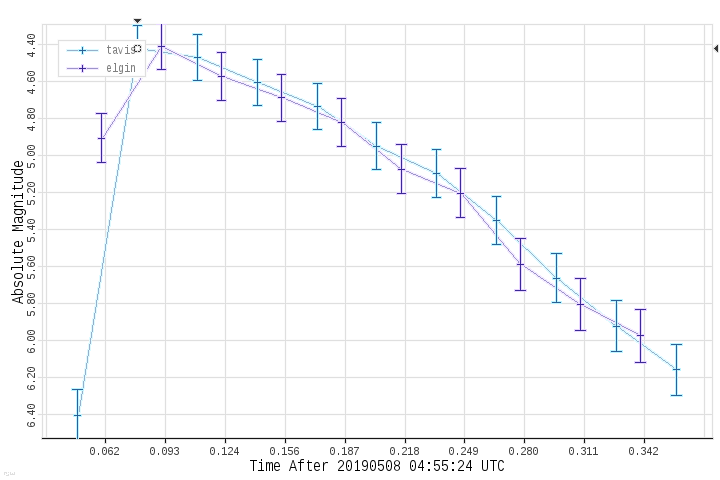}
         \caption{20190508:045524}
         \label{fig:20170703:080601}
     \end{subfigure}
     \end{figure}
       
     \begin{figure}[t!]
     \centering
     \captionsetup[subfigure]{labelformat=empty}
    \begin{subfigure}[b]{0.48\textwidth}
         \centering
         \includegraphics[width=\textwidth]{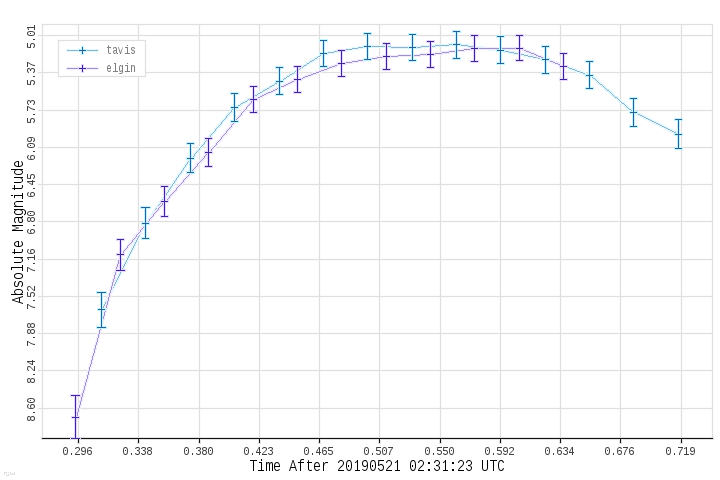}
         \caption{20190521:023123}
         \label{fig:20170702:054951}
     \end{subfigure}
     \begin{subfigure}[b]{0.48\textwidth}
         \centering
         \includegraphics[width=\textwidth]{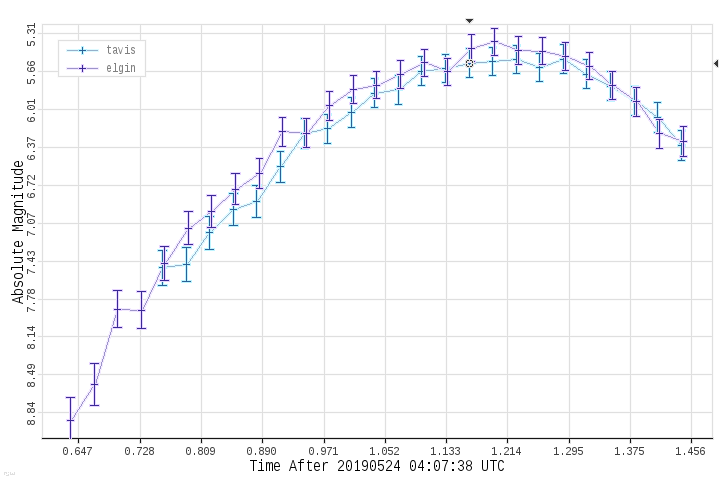}
         \caption{20190524:040738}
         \label{fig:20170703:080601}
     \end{subfigure}
         \begin{subfigure}[b]{0.48\textwidth}
         \centering
         \includegraphics[width=\textwidth]{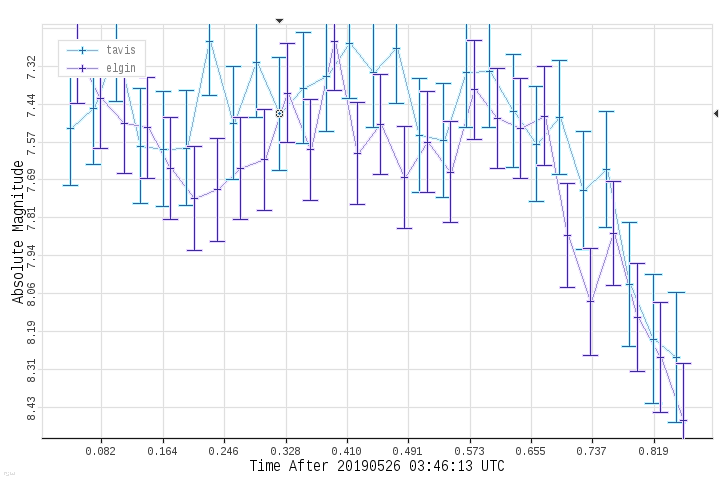}
         \caption{20190526:034613}
         \label{fig:20170702:054951}
     \end{subfigure}
     \begin{subfigure}[b]{0.48\textwidth}
         \centering
         \includegraphics[width=\textwidth]{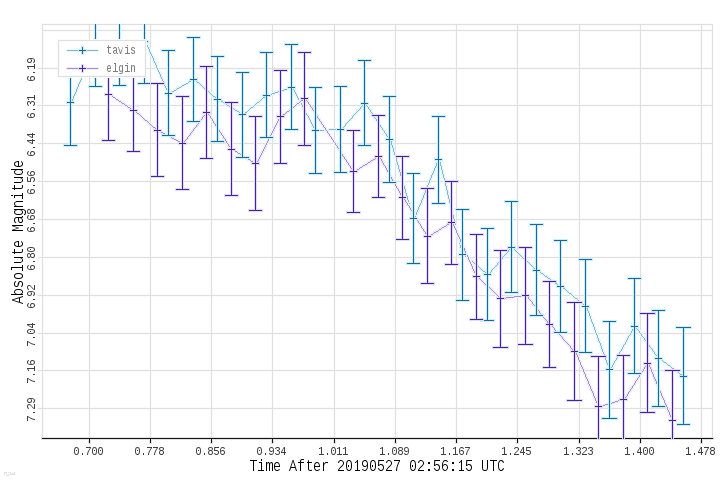}
         \caption{20190527:025615}
         \label{fig:20170703:080601}
     \end{subfigure}
          \begin{subfigure}[b]{0.48\textwidth}
         \centering
         \includegraphics[width=\textwidth]{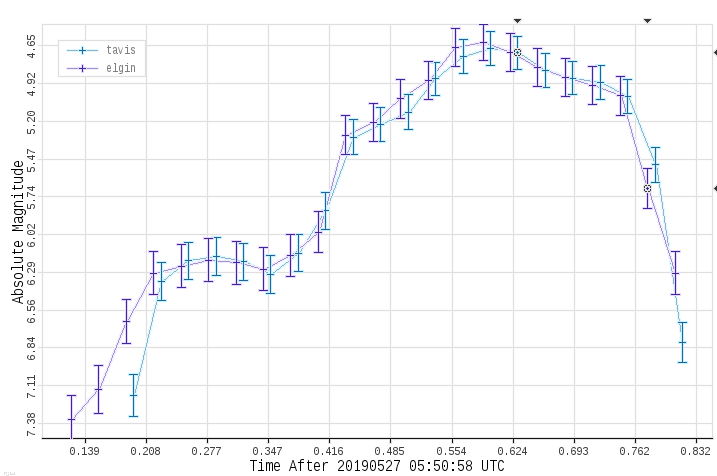}
         \caption{20190527:055058}
         \label{fig:20170702:054951}
     \end{subfigure}
     \begin{subfigure}[b]{0.48\textwidth}
         \centering
         \includegraphics[width=\textwidth]{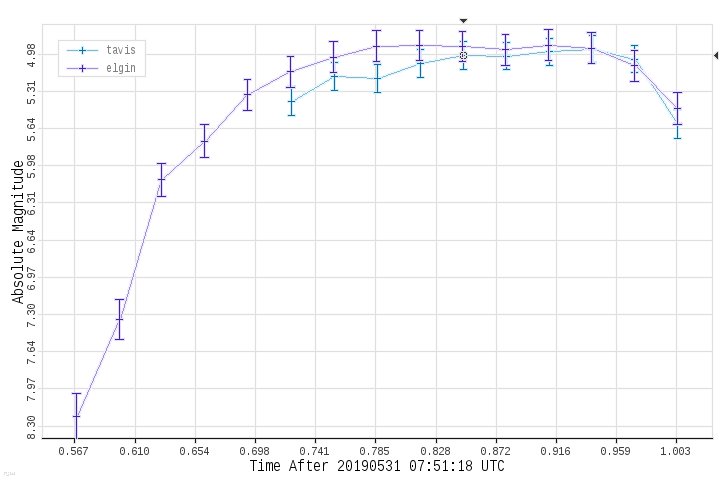}
         \caption{20190531:075118}
         \label{fig:20170703:080601}
     \end{subfigure}
     \end{figure}
     
\end{document}